\documentclass[12pt]{article}

\usepackage{amsmath, amssymb, amsthm}
\usepackage{graphicx}
\usepackage{mathabx}
\usepackage[hidelinks]{hyperref}
\usepackage[margin=3cm]{geometry}
\usepackage{setspace}
\usepackage{subcaption}
\usepackage[natbibapa]{apacite}
\theoremstyle{plain}
\newtheorem{thm}{Theorem}

\newtheorem{assumption}{Assumption}
\newtheorem{definition}{Definition}
\theoremstyle{remark}
\newtheorem{remark}{Remark}

\usepackage{multirow}
\usepackage{booktabs}

\title{Estimating interaction effects with panel data%
    \thanks{
        This paper benefited from discussions at presentations at the Universities of Bristol, G\"ottingen, Groningen, Mainz, WU Vienna, the World Bank, the European Commission's Joint Research Centre, and the Southern Economic Association meetings. We would like to thank the respective participants for their feedback and inputs; comments from Kirill Borusyak, Irene Botosaru, Peter Egger, Michaela Kesina, Adam Lavecchia, Krishna Pendakur, James Powell, Nathanael Vellekoop, Frank Windmeijer, and Jonathan Zhang further helped to significantly improve our paper. We are additionally grateful to Mathieu Couttenier, Mathias Thoenig, and to the Swiss BfS for providing data we used for an empirical application that entered an earlier working paper version of our paper \citep{MurisWacker2022}, and to Eline Koopman for excellent research assistance.
    }
}

\author{
    Chris Muris%
    \thanks{Department of Economics, McMaster University,
            \href{mailto:muerisc@mcmaster.ca}{muerisc@mcmaster.ca}
    } \and 
    Konstantin M. Wacker%
    \thanks{Department of Global Economics \& Management,
            University of Groningen, 
            \href{mailto:k.m.wacker@rug.nl}{k.m.wacker@rug.nl}
    }
    \thanks{GLO Fellow (Global Labor Organization)}
}

\date{\today}

\begin{document}

\onehalfspacing
\maketitle

\vspace*{-3mm}
\begin{abstract}

This paper analyzes how interaction effects can be consistently estimated under economically plausible assumptions in linear panel models with a fixed $T$-dimension. 
We advocate for a \emph{correlated interaction term estimator} (CITE) and show that it is consistent under conditions that are not sufficient for consistency of the interaction term estimator that is most common in applied econometric work. 
Our paper discusses the empirical content of these conditions, shows that standard inference procedures can be applied to CITE, and analyzes consistency, relative efficiency, inference, and their finite sample properties in a simulation study. 
In an empirical application, we test whether labor displacement effects of robots are stronger in countries at higher income levels. The results are in line with our theoretical and simulation results and indicate that standard interaction term estimation underestimates the importance of a country's income level in the relationship between robots and employment and may prematurely reject a null hypothesis about interaction effects in the presence of misspecification.

\end{abstract}

\newpage

\section{Introduction}

In many econometric applications, the relationship between an outcome variable $Y$ and a regressor $X$ depends on a so-called \emph{interaction variable} $H$. For example, an earlier macroeconomic literature postulated that the effect of foreign capital inflows $X$ on economic growth $Y$ depends on some host-country fundamentals $H$, such as institutional quality, financial development, or a sufficiently educated labor force.\footnote{See, for example, \cite{borenszteinHowDoesForeign1998, alfaroFDIEconomicGrowth2004, durham04, alfaroDoesForeignDirect2010, burnsideAidPoliciesGrowth2000}. Economically, this literature assumes that foreign capital $X$ and interaction variables $H$ complement each other in the production of output $Y$.} In micro data, the effect of new technologies on firms' productivity may depend on firm size, reflecting scale effects. And how a worker responds to a labor market reform may depend on age, reflecting that older individuals value leisure higher and income less in their labor supply choice than younger workers.

Panel data have at least two dimensions, $i$ and $t$, which provides us with opportunities to estimate interaction effects that uni-dimensional data do not have. In particular, panel data give us more flexibility to model heterogeneities in the relationship between $Y$ and $X$, on top of variation in $H$. Yet, those opportunities are rarely exploited in applied econometric analysis, which possibly reflects that theoretical work to date has largely ignored the benefits of modeling heterogeneities across panel units for the estimation of interaction effects in panel data.

Consider the relationship between employment $Y$ and robotization $X$. We may hypothesize that this relationship is more negative in countries $i$ with a higher wage level $H_i$ because firms experience higher pressure to save labor costs. The standard approach to estimating such an interaction effect in a linear model with panel data is to regress employment $Y_{it}$ on robotization $X_{it}$, and its interaction with countries' wage levels $H_i$, together unit-specific fixed effects $\alpha_i$ and possibly additional variables in $Z_{it}$, where $t$ may index time, industries, firms etc.:
\begin{equation} \label{eq:ITEreduced} Y_{it} = \alpha_i + X_{it} \beta  + (X_{it} H_i)\kappa + Z_{it}^\prime \gamma + \nu_{it}.\end{equation}
The conventional approach to estimate interaction effects in a linear model with panel data seems to be least squares based on equation \eqref{eq:ITEreduced}.\footnote{See, for example, \cite{burnsideAidPoliciesGrowth2000, Shambaugh2004, ListStrum2006, AmitiKonings2007, spilimbergoDemocracyForeignEducation2009, EpifaniGancia2009, dufloPeerEffectsTeacher2011, BloomSadunReenen2012, BermanMartinMayer2012, BloomDracaReenen2016, Storeygard2016, AlsanGoldin2019, HerreraTrebesch2020, manacordaLiberationTechnologyMobile2020}.} We refer to this approach as the \emph{interaction term estimator} (ITE). The partial effect of $X$ (robots) on $Y$ (employment) linearly depends on $H$ (wage level) in this model, as is easily verified by taking the partial derivative $\partial{Y}/\partial{X} = \beta + H\kappa$.

Our paper focuses on the interaction effect $\kappa$ and how it can be consistently estimated under economically plausible assumptions with panel data that have a short $T$ dimension. 
Standard linear regression assumptions in application to equation \eqref{eq:ITEreduced} require the error term $\nu$ to be independent of the regressors $X, XH, Z$. 
This assumption is easily violated in many interaction models: the economic reasons for effect heterogeneity in the relationship between $X$ and $Y$ are extensive and rarely independent of the regressors. Recall the example of robots and employment. In a technology-minded country, firms $t$ may find various ways to substitute workers' tasks in the production process by robots. The partial relationship between robots $X$ and employment $Y$ will hence be more negative in those countries than in less tech-minded countries because a robot can replace more workers in the former. Since this aspect is unmodeled in equation \eqref{eq:ITEreduced}, and tech-mindedness is not directly observable, this heterogeneity across countries will end up in the error term $\nu$. And since tech-mindedness likely gives rise to higher robotization $X$ in the first place, we observe a standard endogeneity bias: $\nu$ is correlated with $X$ and $XH$ in equation \eqref{eq:ITEreduced}.

Panel data models for interaction effects can be deceptive because the inclusion of the additive fixed effects $\alpha_i$ may lull the researcher into a false sense of having controlled for unobserved heterogeneity. This is not the case because in the context of interaction effects we are not concerned about (additive) heterogeneity in levels of $Y$ and $X$, which are indeed absorbed through $\alpha_i$, but about \emph{effect heterogeneity} in the relationship between $X$ and $Y$. Our research shows that a common source of such unobserved effect heterogeneity can be easily addressed with panel data. Our exposition focuses on the simple case where the interaction variable $H$ only varies in the $i$ dimension since this is plausibly the key dimension giving rise to heterogeneity in the partial effect of $X$ on $Y$. An extension to interaction variables that vary in the $it$ dimension and an application to higher-dimensionality panel data is provided in an earlier working paper version of this paper \citep{MurisWacker2022} and follows the same rationale: additive fixed effects do not control for important aspects of heterogeneity.

The key contribution of our paper is to advocate for a different approach to the estimation of interaction effects with panel data and to analyze the exogeneity conditions that are are sufficient for consistent estimation of $\kappa$. We therefore develop a framework of correlated unobserved effect heterogeneity that explicitly includes unit-specific effects $\beta_i$ as parameters in the relationship between $X$ and $Y$ \cite[see esp.][]{Chamberlain1992}. Building on this framework, we propose a \emph{correlated interaction term estimator} (CITE) that obtains an estimate for the interaction term coefficient $\kappa$ by first running the regression $Y_{it} = \alpha_i + X_{it}\beta_i + Z_{it}^\prime\gamma + U_{it}$ and subsequently projecting the unit-specific effects $\widehat{\beta}_i$ onto $H_i$ in a second-step regression.\footnote{Such an approach has occasionally been used in applied work. See particularly \cite{couttenierViolentLegacyConflict2019a} and \cite{macurdyEmpiricalModelLabor1981}. \cite{giesselmannInteractionsFixedEffects2020} discuss CITE, but make a case for a double-demeaned estimator. Alternatively, running unit-by-unit regressions to recover common parameters is a strategy that has been used in different panel data settings, see e.g. \cite{fernandezvalNonadditiveUnobservedHeterogeneity2013a}. In any case, we are not aware of existing work that demonstrates that ITE and CITE are distinct and that analyzes their asymptotic properties.} Since this second-step regression of $\widehat{\beta}_i$ onto $H_i$ is a simple cross-section OLS regression, it is easily to implement and can be assessed with the extensive, well-known statistical toolkit that has been developed for OLS regressions.\footnote{If one is interested in time-varying interaction variables, those can be considered part of $Z$ and the second step can be omitted. Unobserved effect heterogeneity across panel units is directly absorbed in $\widehat{\beta}_i$ in this case. See an earlier working paper version of our paper for a more explicit treatment of such time-varying interactions \citep{MurisWacker2022}.}

A key result of the rigorous comparison of ITE and CITE that we provide in this paper is that the exogeneity restrictions that guarantee the consistency of CITE for $\kappa$ are not sufficient for the consistency of ITE. On top of conditions that apply to both estimators, CITE requires unobserved effect heterogeneity to be uncorrelated with the interaction variable $H$, while ITE requires this effect heterogeneity to be uncorrelated with all right-hand side regressors $H, X, Z$ for consistent estimation of $\kappa$.  This suggests that researchers need strong exogeneity arguments when estimating interaction terms with ITE in panel data.

One plausible concern about our newly proposed CITE is that its superior bias protection comes at the cost of efficiency. By treating the unit-specific effects $\beta_i$ as parameters that need to be estimated, one may expect a higher variability of CITE for $\kappa$, compared to ITE. 
 We present simulation results demonstrating that this is not necessarily the case when considering different Monte Carlo designs. There is hence no obvious penalty for using CITE in terms of efficiency, while its advantages in terms of consistency are clear.

Another advantage of CITE is that we can demonstrate that standard inference procedures, available in common software, can be applied under economically plausible assumptions about the structure of error terms. The finite sample performance of these inference procedures is also demonstrated in our simulation section. Finally, CITE does not require a large number of time periods. Our paper develops theoretical consistency results under fixed-$T$ and our simulations suggest that relative efficiency is often similar to ITE around $T=4$, depending on the simulation design. The fact that we can prove consistency of CITE for fixed-$T$ also sets our paper apart from the literature on heterogeneity in large-$T$ panels.\footnote{\cite{balliInteractionEffectsEconometrics2013} observe that unobserved effect heterogeneity can lead to inconsistency in the ITE, and suggest that ``if the time-series dimension of the data is large, one may directly allow for country-varying slopes''. We show that such slopes should generally be preferred regardless of the time-series dimension.} 

\subsection*{Related literature}

Our framework builds on the literature on correlated random coefficient models \cite[see][]{Chamberlain1992,arellanoIdentifyingDistributionalCharacteristics2012,grahamIdentificationEstimationAverage2012,laage2024crcpublished,sasakiSlowMoversPanel2021a}. Specifically, we build on Corollary 1 in \cite{arellanoIdentifyingDistributionalCharacteristics2012} to study the estimation of interaction effects in panel models. We adapt and extend the theoretical results in this literature to advocate for CITE over ITE.

Models with correlated coefficient heterogeneity have previously been used to study the identification and estimation of average treatment effects in various settings, see for example
\cite{wooldridgeFixedEffectsRelatedEstimators2005}, \cite{murtazashviliFixedEffectsInstrumental2008},
\cite{verdierAverageTreatmentEffects2020,dechaisemartinTwoWayFixedEffects2022a},
\cite{sloczynskiInterpretingOLSEstimands2022}, 
\cite{chaisemartinMoreRobustEstimators2023}, 
\cite{winkelmannNeglectedHeterogeneitySimpsons2024},
\cite{dechaisemartinCredibleAnswersHard2024}.
This literature is mostly concerned with the identification and estimation of (convex combinations of) average coefficients, partial effects, and treatment effects, from cross-sectional and panel data. 
In contrast, we focus on the estimation of \emph{interaction effects} from panel data in the presence of unobserved effect heterogeneity.

\subsection*{Organization and preview from an applied perspective}

Section~\ref{sec:framework} introduces the unobservable effect heterogeneity model.
Section~\ref{sec:estimation} formally defines the two estimators.
Section~\ref{sec:additional_results} presents our main results about the consistency of CITE.
Section~\ref{sec:variances} discusses inference.
Section~\ref{sec:monte-carlo} contains a Monte Carlo simulation study.
Section~\ref{sec:application} contains an empirical application where we investigate whether worker displacement effects of robots are stronger in countries at higher income levels. 
Section~\ref{sec:conclusion} concludes and discusses some avenues for future research. 

In sections \ref{sec:framework} - \ref{sec:variances}, which focus on theory, we have mostly opted for a notation of within-transformed stacked variables to facilitate the theoretical exposition. Applied readers who are at less comfort with this notation, or less interested in the details of our theoretical results, are suggested to focus on the following aspects: the part of section~\ref{sec:framework} prior to notational definitions to understand the unobservable effect heterogeneity model we have in mind and the first two paragraphs of section~\ref{sec:estimation} for an informal definition of ITE and CITE. 

The crucial exogeneity conditions required for consistent estimation of interaction terms in panel data can be found in Assumptions \ref{assu:Exogeneity-2}-\ref{assu:Exogeneity-1-strengthened} in Section \ref{sec:additional_results}. It is particularly instructive to compare Assumption \ref{assu:Exogeneity-1}, which is sufficient for consistency of CITE, to the stronger Assumption \ref{assu:Exogeneity-1-strengthened}, which is required for consistency of ITE. Remark \ref{rem:assumpt_comp} compares both assumptions with an example from our application.

The essence of section~\ref{sec:variances} for applied researchers is that conventional heteroskedasticity-robust standard errors in the second-stage projection of $\widehat{\beta}_i$ onto $H_i$ are sufficient for valid inference under CITE. Section ~\ref{sec:monte-carlo} is accessible without delving into technical or notational details, but can also be skipped if one is already convinced about the advantages of CITE in terms of consistency, relative efficiency, inference, and their finite sample properties. The application section ~\ref{sec:application} will be most instructive for applied readers to understand the implementation of our approach in practice.

\section{Model} \label{sec:framework}

We are interested in estimating \textit{interaction effects} in static linear panel models with a short $T$ dimension. 
In our framework, the effect of $X$ on $Y$ linearly depends on observables $H$ and on additional \emph{unobservable} sources of effect heterogeneity. 
We model this unobserved heterogeneity in the effect of $X$ on $Y$ as follows:
\begin{align}
Y_{it}    &= \alpha_i + X_{it}\beta_{i} + Z_{it}^\prime\gamma + U_{it},
          &i= 1,\cdots,n, \quad t = 1,\cdots,T,
  \label{eq:outcome-equation} \\
\beta_{i} &= \kappa_0 + \kappa_1 H_{1i} + \kappa_2 H_{2i} + ... + \epsilon_{i} = H_{i}^\prime\kappa + \epsilon_{i},
          &i= 1,\cdots,n,
  \label{eq:heterogeneity-equation}
\end{align}
where $i$ indexes cross-sectional units, and $t$ indexes time periods (or other panel dimensions). The beginning of Section \ref{sec:estimation} clarifies how the standard panel interaction term model relates to this framework.

The \emph{outcome equation}~\eqref{eq:outcome-equation} of our framework describes how a dependent variable $Y_{it} \in \mathbb{R}$ responds to a change in a regressor $X_{it} \in \mathbb{R}$,%
\footnote{Our analysis is easily generalized to vector-valued $X_{it}$.}
allowing for additive fixed effects $\alpha_i \in \mathbb R$, control variables $Z_{it} \in \mathbb{R}^{K_z}$, and an error term $U_{it} \in \mathbb R$. 
It is important to note that the control variables $Z_{it}$ may include interaction terms with $it$ variation. As an earlier working paper version of our paper shows more thoroughly, CITE poses less restrictive exogeneity conditions for their consistency as well \citep{MurisWacker2022}.

The \emph{heterogeneity equation}~\eqref{eq:heterogeneity-equation} of our framework relates effect heterogeneity $\beta_i$ to observed, time-invariant interaction variables $H_i \in \mathbb{R}^{K_h}$. The associated interaction term coefficient $\kappa \in \mathbb{R}^{K_h}$ is the main object of interest in our paper.\footnote{Note that we obtain the constant $\kappa_0$ in the heterogeneity equation \eqref{eq:heterogeneity-equation} by setting the first element of all $H_i$ equal to 1 and that this constant can be interpreted as the average $\beta_i$ across panel units, conditional on all other elements of $H_i$.} The effect of $X$ on $Y$ may additionally vary, even when holding $H_i$ constant, because of \emph{unobserved effect heterogeneity} $\epsilon_i \in \mathbb R$. 
The relationship between unobserved effect heterogeneity and $(X,H)$ plays an important role in our analysis (in a similar way as the relationship between unobserved heterogeneity $\alpha_i$ and $X$ is important in the conventional additive fixed effect model).

\begin{remark}
    The index $t$ need not refer to time, but can refer to students $t$ for a given classroom $i$, counties $t$ within a given state $i$, employees $t$ within a given firm $i$, etc. The empirical application in Section \ref{sec:application} provides an example and section 6.2 in an earlier working paper version illustrates how our framework applies to higher-dimensional panel data \citep{MurisWacker2022}.
\end{remark}

Some additional notation simplifies the exposition in the remainder of the paper. 
First, estimation will be based on the within-transformed outcome equation:
\begin{equation}
    \widetilde Y_{it}  = \widetilde X_{it}\beta_{i} + \widetilde Z_{it}^\prime\gamma + \widetilde U_{it},
    \label{eq:within-transformed-outcome-equation}
\end{equation}
where
\[ 
    \widetilde{Y}_{it} = Y_{it} - \frac{1}{T} \sum_{s=1}^{T} Y_{is}, \quad
    \widetilde{X}_{it} = X_{it} - \frac{1}{T} \sum_{s=1}^{T} X_{is},
\]
and $\widetilde Z_{it}$ and $\widetilde U_{it}$ are defined analogously.

Second, the reduced form of the within-transformed equations \eqref{eq:outcome-equation}--\eqref{eq:heterogeneity-equation} is
\begin{align}
    \widetilde Y_{it} &= \widetilde X_{it} (H_i^\prime \kappa + \epsilon_i) + \widetilde Z_{it}^\prime \gamma + \widetilde U_{it} \nonumber \\
           &= \underbrace{\widetilde X_{it} H_i^\prime}_{\Psi_{it}^\prime} \kappa + \widetilde Z_{it}^\prime \gamma + 
              \underbrace{\left(\widetilde U_{it} + \widetilde X_{it} \epsilon_i\right)}_{V_{it}},
    \label{eq:simple_reduced_form}
\end{align}
with an error term $V_{it}$ that depends on the unobserved effect heterogeneity $\epsilon_i$. 

Third, we collect the within-transformed outcome equation \eqref{eq:within-transformed-outcome-equation} across $t=1,\cdots,T$ for a given $i$ to obtain
\begin{equation}
\widetilde Y_{i}= \widetilde X_{i} \beta_{i} + \widetilde Z_{i}\gamma + \widetilde U_{i},
 \label{eq:outcome_heterogeneous_coefficients_stacked}
\end{equation}
where
\[
    \widetilde{X}_i = (\widetilde{X}_{i1},\cdots,\widetilde X_{iT}) \in \mathbb R^T, 
\]
and $\widetilde Y_{i} \in \mathbb R^T$, $\widetilde X_{i} $, $\widetilde Z_{i} \in \mathbb R^{T \times K_z}$, and $\widetilde U_{i} \in \mathbb R^T$. 
Accordingly, we obtain  
\begin{align}
    \widetilde Y_{i} &= \Psi_{i}\kappa + \widetilde Z_{i}\gamma + V_i, 
    \label{eq:transformed_and_stacked}
    \end{align}
where $\Psi_i = \widetilde X_i H_i^\prime \in \mathbb R^{T \times K_h}$ and $V_i = \widetilde U_i + \widetilde X_i \epsilon_i \in \mathbb R^T$ are stacked versions of $\Psi_{it}$ and $V_{it}$, respectively.

\section{Estimators}
\label{sec:estimation}

We consider two estimators for the interaction term coefficient $\kappa$: the \emph{interaction term estimator} (ITE) 
and the \emph{correlated interaction term estimator} (CITE).
The ITE is a fixed effects regression%
\footnote{By ``fixed effects regression'' we mean a regression that includes a dummy variable for each $i$, i.e. the within estimator.}
of $Y$ on the interaction term $\Psi = XH$ and $Z$, as suggested by the reduced form~\eqref{eq:simple_reduced_form}.
It is the default approach in applied research when the effect of $X$ is expected to depend on $H$: \emph{``just add an interaction term''}. This standard ITE panel interaction model is usually written along the lines $Y_{it} = \alpha_i + X_{it} \beta + X_{it} H_{1i} \kappa_1 + Z'_{it}\gamma + \nu_{it}$ in many applied papers, which is a special case of our framework that one obtains by substituting equation \eqref{eq:heterogeneity-equation} into equation \eqref{eq:outcome-equation}, imposing $\epsilon_i = 0 \ \forall \ i$, and setting $\beta = \kappa_0$. A formal definition of ITE is provided in Section~\ref{sec:ITE}.

The CITE, defined formally in Section~\ref{sec:DVAITE}, is a two-step estimator.
The first step is a fixed effects regression of $Y$ on $Z$ and interactions of $X$ with dummy variables for each $i$: $Y_{it} = \alpha_i + X_{it}\beta_i + Z'_{it}\gamma + U_{it}$.
The coefficients on the dummy variable interactions, $\widehat \beta_i$, are individual-specific effects of $X$ on $Y$ that capture unobserved effect heterogeneity. 
The second step is a regression of $\widehat \beta_i$ on $H$.
CITE hence directly follows the structure in equations \eqref{eq:outcome-equation}--\eqref{eq:heterogeneity-equation}: the first-step regression is based on the outcome equation~\eqref{eq:outcome-equation}, and the second-step regression is based on heterogeneity equation~\eqref{eq:heterogeneity-equation}. Practical implementation of CITE in standard software is simple: it just requires interacting $X$ with the unit-specific dummy variables, saving the estimated coefficients, and regressing them on $H$ in a second step.

The remainder of this manuscript assumes a random sample is available.
\begin{assumption}[Random sampling]
\label{assu:sampling}
For each $i=1,\cdots,n$, the observed data is $$W_i = \left(Y_{i1},\cdots,Y_{iT},X_{i1},\cdots,X_{iT},Z_{i1},\cdots,Z_{iT}, H_i\right),$$ generated by equations \eqref{eq:outcome-equation}--\eqref{eq:heterogeneity-equation}. 
The random sequence $\{W_i\}_{i=1}^n$ is independent and identically distributed.
\end{assumption}

\subsection{Correlated interaction term estimator}
\label{sec:DVAITE}

Sufficient variation in $\widetilde X_{i}$ for all $i$ is required for the CITE to be well-defined. 
\begin{assumption}
\label{assu:switchers}
There exists an $h>0$ such that
\[
    \inf_{i}
    \left(
        \sum_{t=1}^T \widetilde X_{it}^2 
    \right)
    \geq h.
\]
\end{assumption}
This guarantees that the least squares estimators for the $(\beta_i)$ are well-defined,
which avoids the identification issues addressed in~\cite{grahamIdentificationEstimationAverage2012} and~\cite{arellanoIdentifyingDistributionalCharacteristics2012}. 
If Assumption \ref{assu:switchers} does not hold for all $i$, 
our analysis applies to the subpopulation for which it does hold.

To obtain the within-transformed data, we define the residual maker matrix
\[ 
    M_{i} = 
        I_T - 
        \widetilde X_{i}
        \left(
            \widetilde X_{i}^\prime \widetilde X_{i}
        \right)^{-1}
            \widetilde X_{i}^\prime, 
\]
where $I_T$ is the $T \times T$ identity matrix.
Premultiplication by $M_i$ obtains
\begin{align}
M_{i}\widetilde Y_{i} &= M_{i} \widetilde X_{i} \beta_{i} +
              M_{i} \widetilde Z_{i}\gamma + 
              M_{i} \widetilde U_{i}\nonumber \\
           &= M_{i} \widetilde Z_{i}\gamma + M_{i} \widetilde U_{i},
           \label{eq:transformed_and_stacked-DVAITE}
\end{align}
and estimation of $\gamma$ can be based on equation  \eqref{eq:transformed_and_stacked-DVAITE}.
\begin{assumption}
\label{assu:rank-transformed}
The matrix
$E\left(\widetilde Z_{i}^\prime M_{i} \widetilde Z_{i}\right)$
is invertible.
\end{assumption}
Assumptions~\ref{assu:sampling}--\ref{assu:rank-transformed} guarantee that the following is well-defined for large $n$.
\begin{definition}
\label{def:DVAITE_theta}
CITE for $\gamma$ is given by
\[
    \widehat{\gamma}_{n} 
    =
    \left(
        \sum_{i=1}^n \widetilde Z_{i}^\prime M_{i} \widetilde Z_{i}
    \right)^{-1} 
        \sum_{i=1}^n \widetilde Z_{i}^\prime M_{i} \widetilde Y_{i}.
\]
\end{definition}
To define the CITE for $\kappa$, we also need sufficient variation in $H_i$.
\begin{assumption}
\label{assu:Full-rank-H}
The matrix $E\left(H_{i}^\prime  H_{i}\right)$ is invertible.
\end{assumption}
For each $i$, the estimator for the individual-specific effect of $X$ on $Y$ is
\begin{align*}
    \widehat \beta_i 
    &=
    \left(\widetilde X_{i}^\prime \widetilde X_{i}\right)^{-1}
    X_{i}^\prime \left(Y_{i}-Z_{i}\widehat{\gamma}_n\right).
\end{align*}
The CITE for $\kappa$ is obtained from a regression of $\widehat \beta_i$ on $H_i$.
\begin{definition}
\label{def:DVAITE_kappa}
The CITE for $\kappa$ is
\[
    \widehat{\kappa}_n =
    \left(\sum_{i=1}^n H_{i}^\prime H_{i}\right)^{-1} 
          \sum_{i=1}^n H_{i}^\prime \widehat{\beta}_i.
\]
\end{definition}

\subsection{Interaction term estimator} \label{sec:ITE}

The ITE is based on equation \eqref{eq:transformed_and_stacked}.
To formally define it, we project out the fixed effects $\alpha_i$ using the within transformation.
For the ITE to be well-defined, we require the following `no multicollinearity' condition:
\begin{assumption}
\label{assu:rank-transformed-ITE}
The matrix 
$E\left(
    [\Psi_{i} \; \widetilde Z_i]^\prime
    [\Psi_{i} \; \widetilde Z_i]
\right)$ 
is positive definite.
\end{assumption}
\begin{definition}
\label{def:ITE}
Given Assumption \ref{assu:rank-transformed-ITE}, 
the ITE for $\theta = (\kappa,\gamma)$ is well-defined:
\begin{align*}
    \widecheck{\theta}_{n} 
    &=
    \left(\sum_{i=1}^n
        [{\Psi}_{i} \; \widetilde Z_i]^\prime
        [{\Psi}_{i} \; \widetilde Z_i]
    \right)^{-1}
    \sum_{i=1}^n
        [{\Psi}_{i} \; \widetilde Z_i]^\prime
         \widetilde{Y}_{i} \\
    &= 
    (\widecheck \kappa_n, \widecheck \gamma_n).
\end{align*}
\end{definition}

\section{Consistency}
\label{sec:additional_results}

We now show that a set of exogeneity conditions that is sufficient for consistency of CITE is not sufficient for consistency of ITE.
Throughout, we maintain the following strict exogeneity assumption.
\begin{assumption}[Strict exogeneity]
    \label{assu:Exogeneity-2}$E\left(\left.U_{i}\right|X_{i},Z_{i},H_{i}\right) = 0$.
\end{assumption}
This assumption is similar to the strict exogeneity assumption that is standard in the literature on correlated random coefficient panel models,\footnote{See \cite{Chamberlain1992, arellanoIdentifyingDistributionalCharacteristics2012, grahamIdentificationEstimationAverage2012}. See \cite{laageCorrelatedRandomCoefficient2020b} for an approach without exogeneous regressors.} and in textbook treatments of fixed-$T$ linear panel models with additive fixed effects. 
If one is only interested in $\gamma$, and not in $\kappa$, Assumption \ref{assu:Exogeneity-2} can be relaxed for CITE (but not for ITE) to
$E\left(\left.U_{i}\right|X_{i},Z_{i}\right) = 0.$ 

\subsection{CITE}\label{sec:correct_CITE}

It is sufficient for the consistency of CITE that the unobserved effect heterogeneity $\epsilon_i$ is orthogonal to the time-invariant interaction variable $H_i$.
\begin{assumption}[Exogeneity, $\epsilon$]$E\left(\left.\epsilon_{i}\right|H_{i}\right)=0.$
\label{assu:Exogeneity-1}
\end{assumption}
This assumption is easily understood when recalling that the second step of CITE is a cross-sectional OLS regression: it is equivalent to the textbook OLS exogeneity condition that requires the regressor $H_i$ to be orthogonal to the error term $\epsilon_i$.

To understand the advantages of CITE in applied econometric work, it is crucial to emphasize that Assumption \ref{assu:Exogeneity-1} does not involve the regressors $\left(X,Z\right)$. We discuss this advantage in contrast with ITE, including an example, in Remark \ref{rem:assumpt_comp} at the end of Section \ref{sec:additional_results}. 
\begin{thm}
\label{thm:consistency-PRC-theta}
(i) If Assumptions~\ref{assu:sampling}--\ref{assu:rank-transformed}~and~\ref{assu:Exogeneity-2} hold, and if
$E\left\|Z_i\right\|^2$
and $E\left\|Y_i\right\|^2$
are bounded,
then
$\widehat{\gamma}_{n}\stackrel{p}{\to}\gamma \text{ as }n\to\infty.$
  (ii) If additionally, Assumptions~\ref{assu:Full-rank-H} and \ref{assu:Exogeneity-1} hold, and if $E\|H_i\|^2$, $E\|\beta_i\|^2$, $E\|\widetilde X_i\|^4$, and $E\|\widetilde Z_i\|^4$ are bounded,
then
$\widehat{\kappa}\stackrel{p}{\to}\kappa \text{ as }n\to\infty.$
\end{thm}
For consistency of CITE, we do not need to restrict the joint distribution of $(Z,X,\epsilon)$ beyond the existence of certain moments.
The proof of consistency is standard and can be found in Appendix \ref{sec:proofs}. 
Because distribution theory for ITE and CITE is standard, we omit the remaining proofs.

\subsection{Inconsistency of ITE}\label{sec:correct_ITE_fail}

Theorem \ref{thm:consistency-PRC-theta} establishes that the exogeneity restrictions in Assumptions \ref{assu:Exogeneity-2} and \ref{assu:Exogeneity-1} guarantee the consistency of CITE for $\kappa$ and $\gamma$. 
We now show that the exogeneity restrictions are not sufficient for the consistency of ITE for $\kappa$. 

Consider the special case that $H$ is scalar, and that there is no $Z$, so the reduced form simplifies to:
\[
    \widetilde Y_{it} 
    = 
    \widetilde X_{it} H_i \kappa + \left(\widetilde U_{it} + \widetilde X_{it} \epsilon_i\right).
\]
The ITE is simply a linear regression of $\widetilde Y$ on $\widetilde X H$.
Consistency of OLS requires that the error term is orthogonal to the regressors.
Given Assumption \ref{assu:Exogeneity-2}, 
\[
    E\left(\widetilde X_{it} H_i \widetilde U_{it}\right) = 0.
\]
However, under the exogeneity Assumptions~\ref{assu:Exogeneity-2} and \ref{assu:Exogeneity-1} we may have 
\[
    E\left(\widetilde X_{it}^2 H_i \epsilon_i\right) \neq 0.
\]
It is straightforward to construct data generating processes where this happens, see our simulation study in Section \ref{sec:monte-carlo}. 

To ensure the consistency of ITE, one could further strengthen the conditions.
For example, strengthening Assumption \ref{assu:Exogeneity-1} to the following stronger \emph{correlated random effects} assumption restores consistency:

\begin{assumption}[Exogeneity, $\epsilon$, strengthened]$E\left(\left.\epsilon_{i}\right|H_{i},X_i,H_i\right)=0.$
\label{assu:Exogeneity-1-strengthened}
\end{assumption}

This restriction requires the unobserved effect heterogeneity $\epsilon_i$ to be orthogonal to $X$ and $Z$ in addition to $H$. Consistency of ITE hence requires that unobserved effect heterogeneity is independent of all relevant regressors in the model.

\begin{remark} \label{rem:assumpt_comp}
Comparison of Assumptions \ref{assu:Exogeneity-1} and \ref{assu:Exogeneity-1-strengthened} facilitates a key insight of our paper: the latter (which ensures consistency of ITE) is more demanding than Assumption \ref{assu:Exogeneity-1} (which is sufficient for consistency of CITE), since it requires additional orthogonality of $\epsilon_i$ to $X$ and $Z$ (in addition to $H$). In the context of our empirical application, where we estimate how the effect of robots $X$ on employment $Y$ depends on interaction variables $H$, ITE requires that heterogeneity in the relationship between robotization and employment across countries is uncorrelated with countries' robot adoption. As discussed in our introduction, this is implausible because certain countries may be more open to new technologies, in which case they are more likely to adopt robots $X$ and find ways to replace workers with robots, which gives rise to effect heterogeneity $\epsilon_i$. CITE does not require such effect heterogeneity to be independent of robotization $X$ for consistent estimation of the interaction effect.
\end{remark}

\section{Inference}\label{sec:variances}

In this section, we discuss how to conduct inference on the interaction effect $\kappa$ for both estimators. 
For both estimators, inference is standard and can be performed using widely available software.
We sketch our arguments below for the case where $Z$ is absent. 
The general case with $Z$ is similar, and it is omitted because the additional notation obscures the argument.

In this paper, we are concerned with inference for the partial effects.
Empirical researchers may also be interested in the distribution of partial effects.
Inference for this object is significantly more challenging, see for example \cite{arellanoIdentifyingDistributionalCharacteristics2012} and \cite{fernandez2022dynamic}.

\subsection{ITE}\label{sec:inference_ITE}

Inference for the ITE follows the standard theory for linear panel data models with clustered errors. In panel data, cluster-robust standard errors account for arbitrary correlation patterns in the error terms within each panel unit while maintaining the assumption of independence across units. Clustered standard errors address the fact that observations from the same panel unit (like multiple years of data from the same country) are typically not independent.

Two features of our model make clustering necessary. First, the within transformation induces correlation in the transformed errors $\widetilde{U}_i$ within each panel unit $i$. Second, the presence of unit-specific unobserved effect heterogeneity $\epsilon_i$ in the composite error term $V_i = \widetilde{U}_i + \widetilde{X}_i \epsilon_i$ creates an additional source of within-unit correlation. However, our random sampling assumption (Assumption \ref{assu:sampling}) ensures independence across panel units, making cluster-robust standard errors appropriate.

The asymptotic variance of the ITE takes the standard sandwich form for clustered data:
\[
\text{var}(\widecheck{\kappa}_n)
=
\left(\sum_i \Psi_i^\prime \Psi_i\right)^{-1}
\left(\sum_i \Psi_i^\prime \text{var}(V_i|\Psi_i) \Psi_i\right)
\left(\sum_i \Psi_i^\prime \Psi_i\right)^{-1}.
\]
This variance can be consistently estimated in the usual way.

Implementation of cluster-robust standard errors is straightforward in standard statistical software. For instance, in current STATA versions, one would use the option \texttt{vce(robust)} for a standard panel command with interactions, like \texttt{xtreg y x c.x\#c.h, fe vce(robust)}. Similar commands are available in R, Python, and other commonly used statistical packages. Our simulation study in Section \ref{sec:monte-carlo} documents the finite-sample performance of this inference approach.

\subsection{CITE}\label{sec:inference_CITE}

Because of the two-step nature of CITE, it may be surprising that heteroskedasticity-robust standard errors in the second-step regression are sufficient for valid inference.
To show that this is the case, recall that 
CITE for $\kappa$ is a regression of the estimated individual-specific coefficients $\widehat \beta_i$ on $H_i$,
\[
    \widehat{\kappa}_n =
    \left(\sum_{i=1}^n H_{i}^\prime H_{i}\right)^{-1} 
          \sum_{i=1}^n H_{i}^\prime \widehat{\beta}_i.
\]
The corresponding regression equation is 
\[
    \widehat \beta_i = \beta_i + (\widehat \beta_i - \beta_i) = H_i^\prime \kappa + \epsilon_i + (\widehat \beta_i - \beta_i),
\]
where the second equality follows from the heterogeneity equation \eqref{eq:heterogeneity-equation}.

Therefore, $\widehat{\kappa}_n$ is based on a linear regression with a composite error term $\epsilon_i + (\widehat \beta_i - \beta_i)$. This composite error term is independent across $i$, and heteroskedastic by construction. Independence follows from two observations: first, Assumption \ref{assu:sampling} implies that $\epsilon_i$ is independent across $i$, and second, the first-step estimation errors 
$\widehat \beta_i = \left(\widetilde X_{i}^\prime \widetilde X_{i}\right)^{-1} X_{i}^\prime Y_{i}$ 
are independent across $i$ because they only depend on $(\widetilde{X}_i, \widetilde{Y}_i)$, which are also independent across $i$ by Assumption \ref{assu:sampling}.
The composite error term is heteroskedastic because the variance of the first-step estimation error $(\widehat \beta_i - \beta_i)$ depends on $\widetilde{X}_i$, even if $\epsilon_i$ is homoskedastic.
Consequently, using heteroskedasticity-robust standard errors (\cite{whiteHeteroskedasticityConsistentCovarianceMatrix1980}) in the second-step regression leads to valid inference.

Implementation is straightforward in standard statistical packages. For example, in STATA, after obtaining the unit-specific coefficients `bhat' via a first-stage regression, one would use \texttt{reg bhat H, robust} to obtain valid standard errors. Our simulation study in Section \ref{sec:monte-carlo} confirms good finite-sample performance of this approach.
When control variables $Z$ are present, the argument remains valid asymptotically, 
because $\gamma$ is estimated at rate $\sqrt{nT}$.
The contribution to sampling variation is negligible compared to that from $\widehat \beta_i$. The latter is estimated using $T$ observations.

\section{Monte Carlo simulations}
\label{sec:monte-carlo}

We use a Monte Carlo simulation to document
the bias in ITE,
the relative efficiency of ITE and CITE,
and the finite sample performance of the inference procedures.

We use a data generating process in which all building blocks follow a normal distribution, are mutually independent, and are independent across $i$.
For the heterogeneity equation, we let
$H_i \sim \mathcal{N}\left(1,1\right)$,
$\epsilon_i \sim \mathcal{N}(0,\sigma_\epsilon)$,
and
$\beta_i = \kappa H_i + \epsilon_i.$
For the outcome equation,
we omit $Z$, and construct $X_{it}$ as follows:
\begin{align*}
  \psi_{it} &\sim \mathcal{N}(0,\sigma_x), \quad \lambda_t \sim \mathcal{N}(0,\sigma_l), \\
  X_{it}    &= (1+\psi_{it}) (1+\Delta \times \epsilon_{it}) (1+\lambda_t).
\end{align*}
The parameter $\Delta$ controls the amount of correlation between $X_{it}$ and $\epsilon_{it}$. Values of $\Delta$ away from zero favor CITE because they create a correlation between $X$ and unobserved effect heterogeneity $\epsilon$ (see Section \ref{sec:correct_ITE_fail}). $\lambda_t$ creates common time trends for all panel units. 
Finally, we generate $U_{it} \sim \mathcal N(0,\sigma_u)$, $\alpha_i \sim \mathcal N(0,\sigma_a)$, and compute
$Y_{it} = \alpha_i + X_{it} \beta_i + U_{it}.$

All $\sigma$ parameters default to 1.
We will vary the number of time periods $T$ and the number of cross-section units $n$, as well as the value of the interaction term coefficient $\kappa$ and other design parameters.

\subsection{The bias in ITE}

\begin{figure}[htbp]
    \begin{subfigure}{0.48\textwidth}
        \centering
        \includegraphics[width=\textwidth]{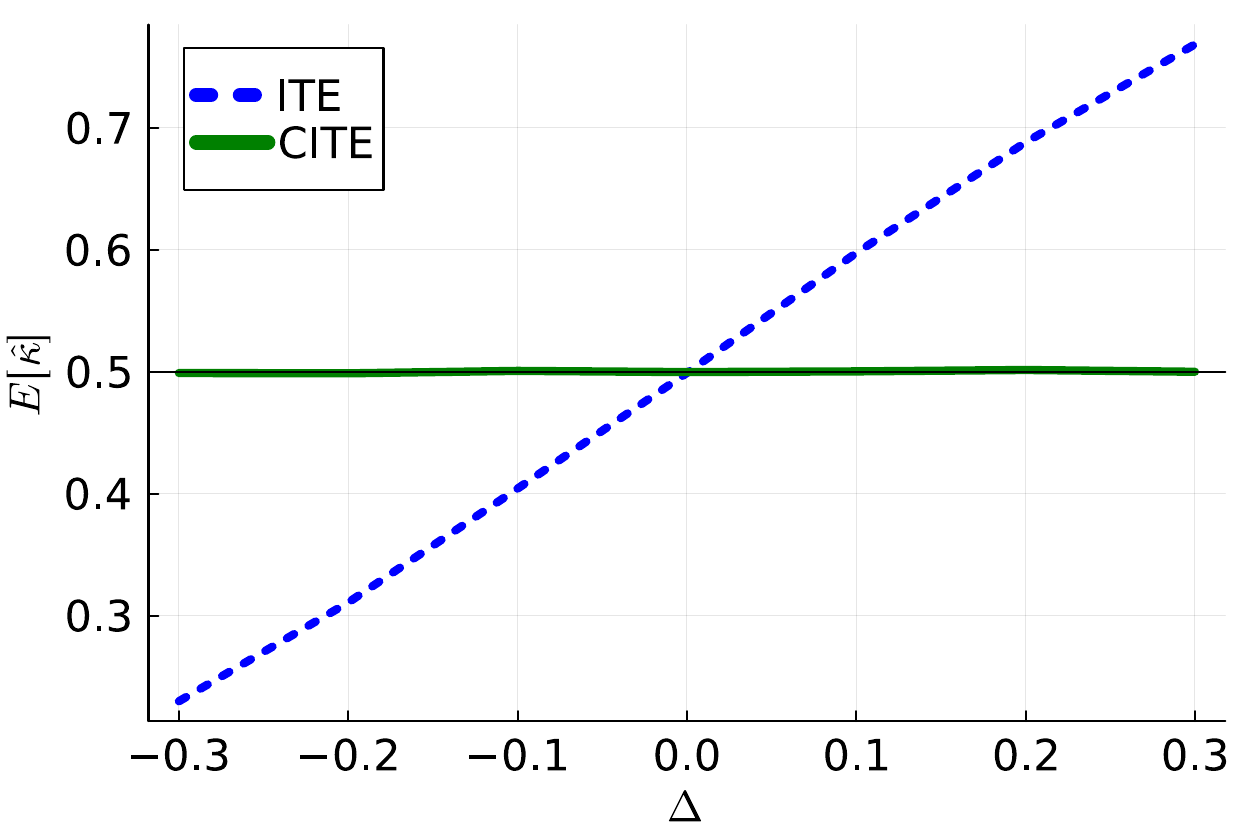}
        \caption{Means as a function of $\Delta$ (true $\kappa$=0.5).}
        \label{fig:bias_delta}
    \end{subfigure}
    \hfill
    \begin{subfigure}{0.48\textwidth}
        \centering
        \includegraphics[width=\textwidth]{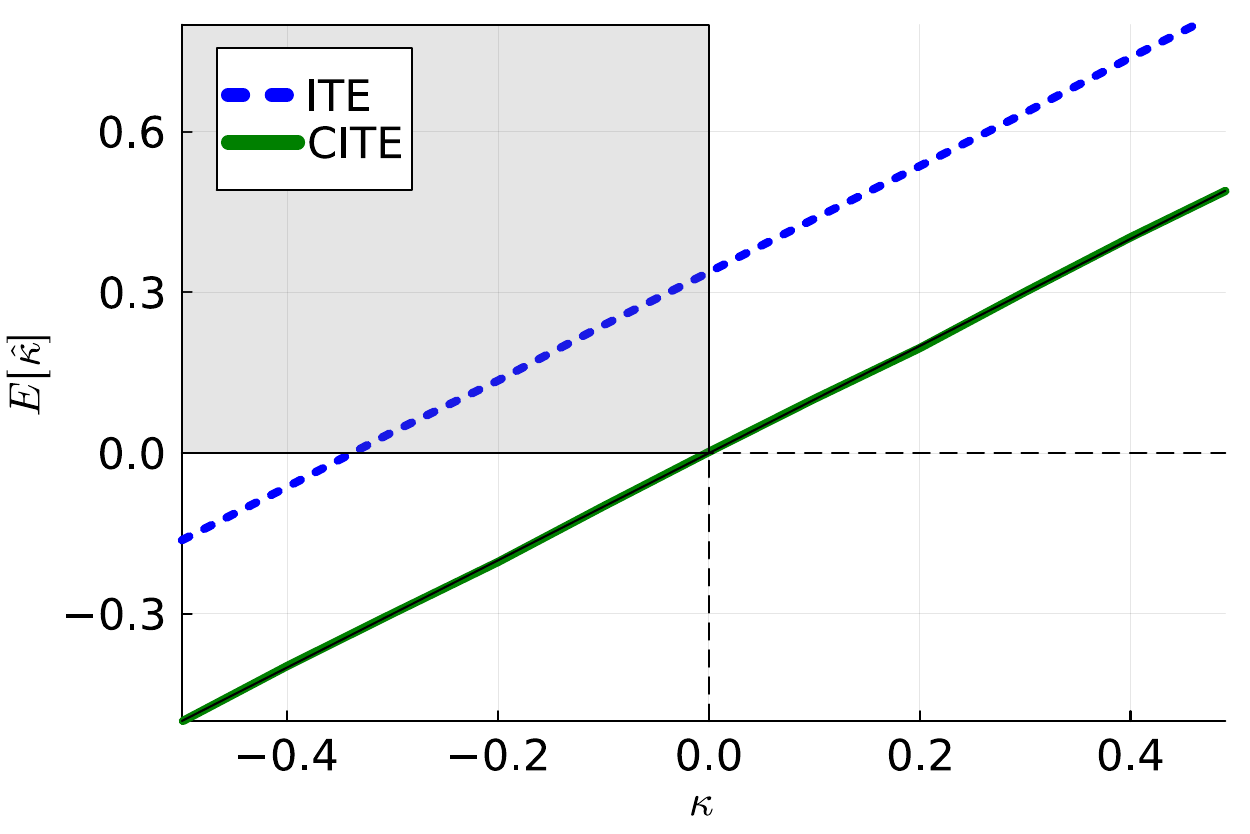}
        \caption{Means as a function of $\kappa$.}
        \label{fig:bias_kappa}
    \end{subfigure}
    \caption{Mean of ITE and CITE.}
    \label{fig:bias}
\end{figure}

We start with the design outlined above, with $n = 100$, $T=5$, and $\kappa = 0.5$. 
Figure \ref{fig:bias_delta} plots the mean of ITE and CITE across 10000 simulations, as a function of $\Delta \in [-0.3,0.3]$. 
At $\Delta = 0$, there is no endogeneity bias in the ITE, and both estimators for $\kappa$ have a mean of 0.5. 
For any $\Delta \neq 0$, the ITE is biased, whereas the mean CITE remains at 0.5.
The bias in the ITE can be severe.

Figure \ref{fig:bias_kappa} plots the mean of ITE and CITE as a function of $\kappa \in [-0.5,0.5]$ at $\Delta = 0.4$.  
An unbiased estimator should follow the 45-degree diagonal, which is achieved by CITE. 
Conversely, the bias in ITE is substantial ($\approx 0.3$) and does not change much when varying $\kappa$.
The shaded area corresponds to values of $\kappa$ where the mean of ITE has the opposite sign of $\kappa$.

\subsection{Relative efficiency}
\label{sec:simulation_efficiency}

Next, we compare the variability of ITE and CITE.
We continue using $n = 100$, set $\Delta = 0$ (otherwise ITE is biased), and set $\kappa = 0$ (results for other values are comparable).
We will vary $T \in \{3,\cdots,8\}$.
In our baseliness results, the $\sigma$ parameters are at $1$. We will consider the effect of changing them on the relative efficiency of ITE and CITE as a function of $T$.

\begin{figure}[htbp]
    \begin{subfigure}{0.48\textwidth}
        \centering
        \includegraphics[width=\textwidth]{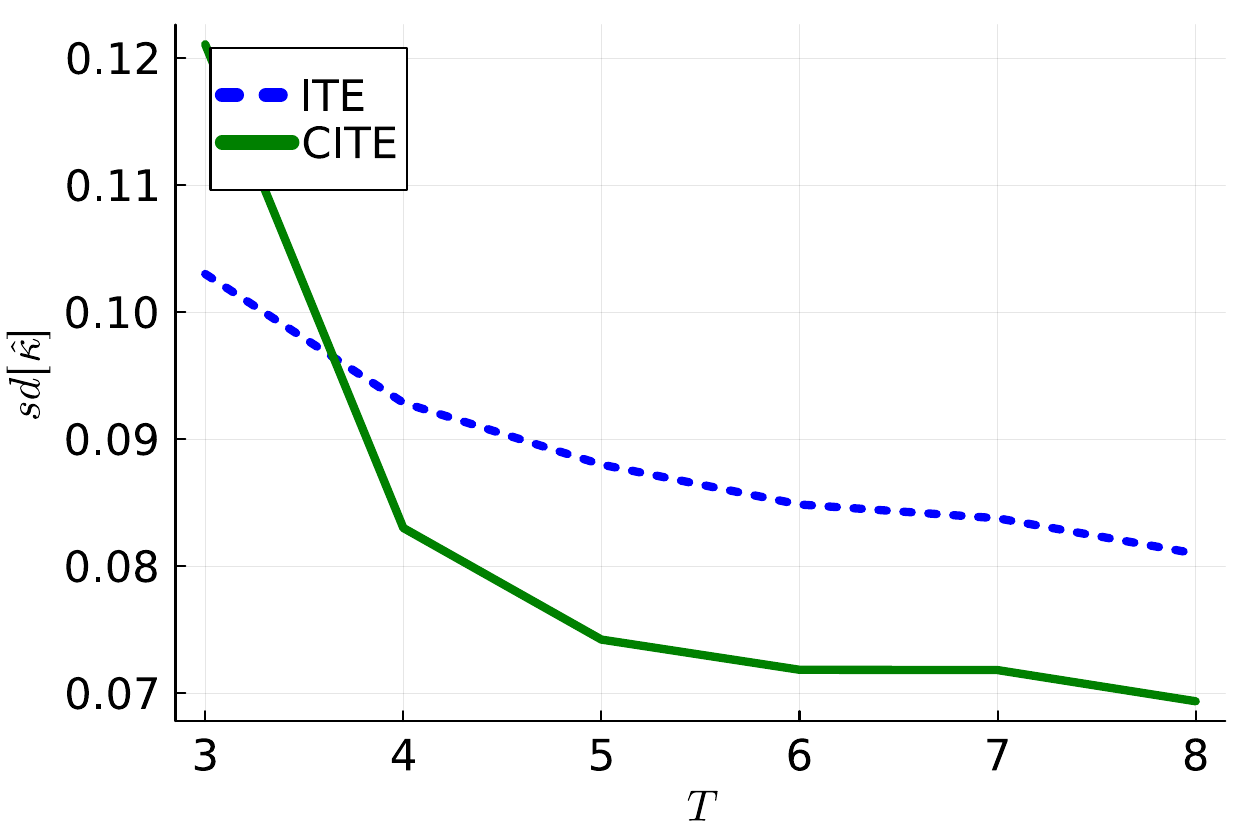}
        \caption{Baseline, all $\sigma = 1$}
        \label{fig:sd_baseline}
    \end{subfigure}
    \hfill
    \begin{subfigure}{0.48\textwidth}
        \centering
        \includegraphics[width=\textwidth]{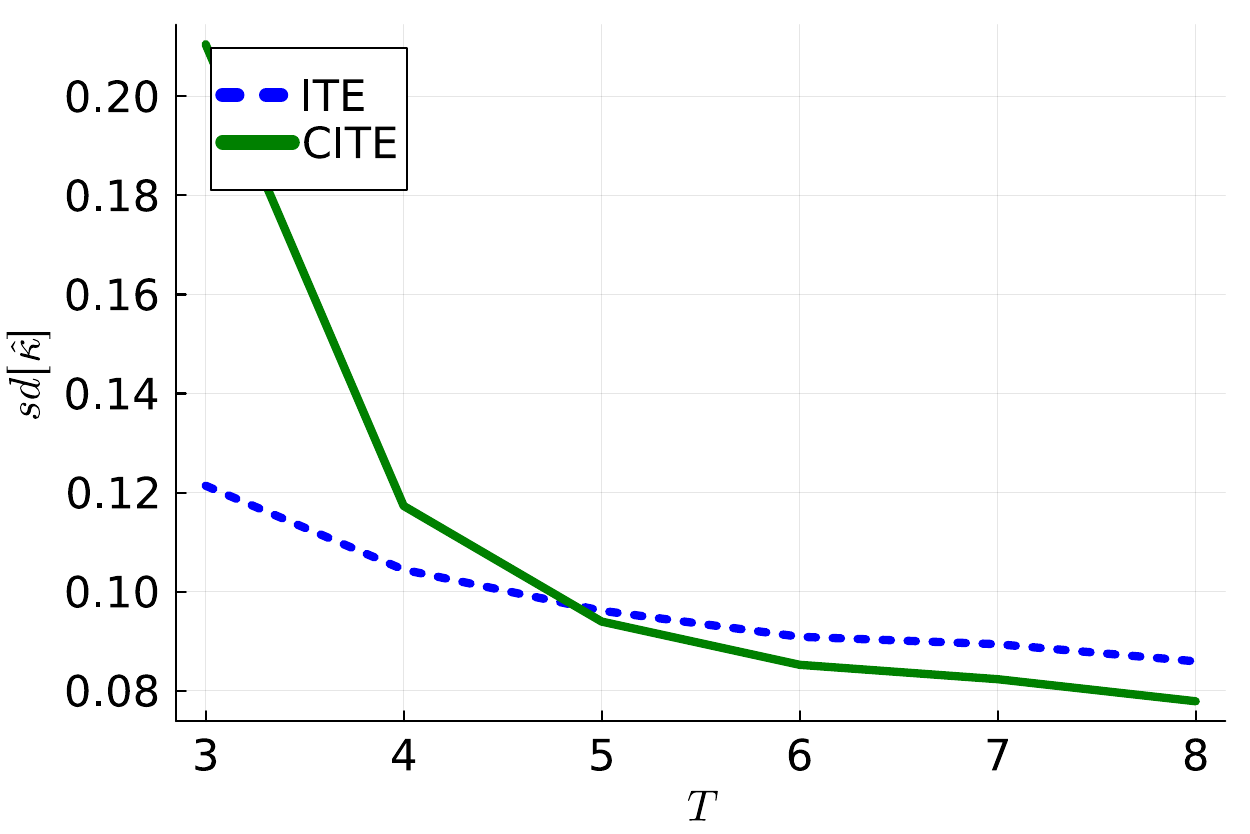}
        \caption{$\sigma_u = 2$}
        \label{fig:sd_u_2}
    \end{subfigure}
    
    \vspace{1em}  
    
    \begin{subfigure}{0.48\textwidth}
        \centering
        \includegraphics[width=\textwidth]{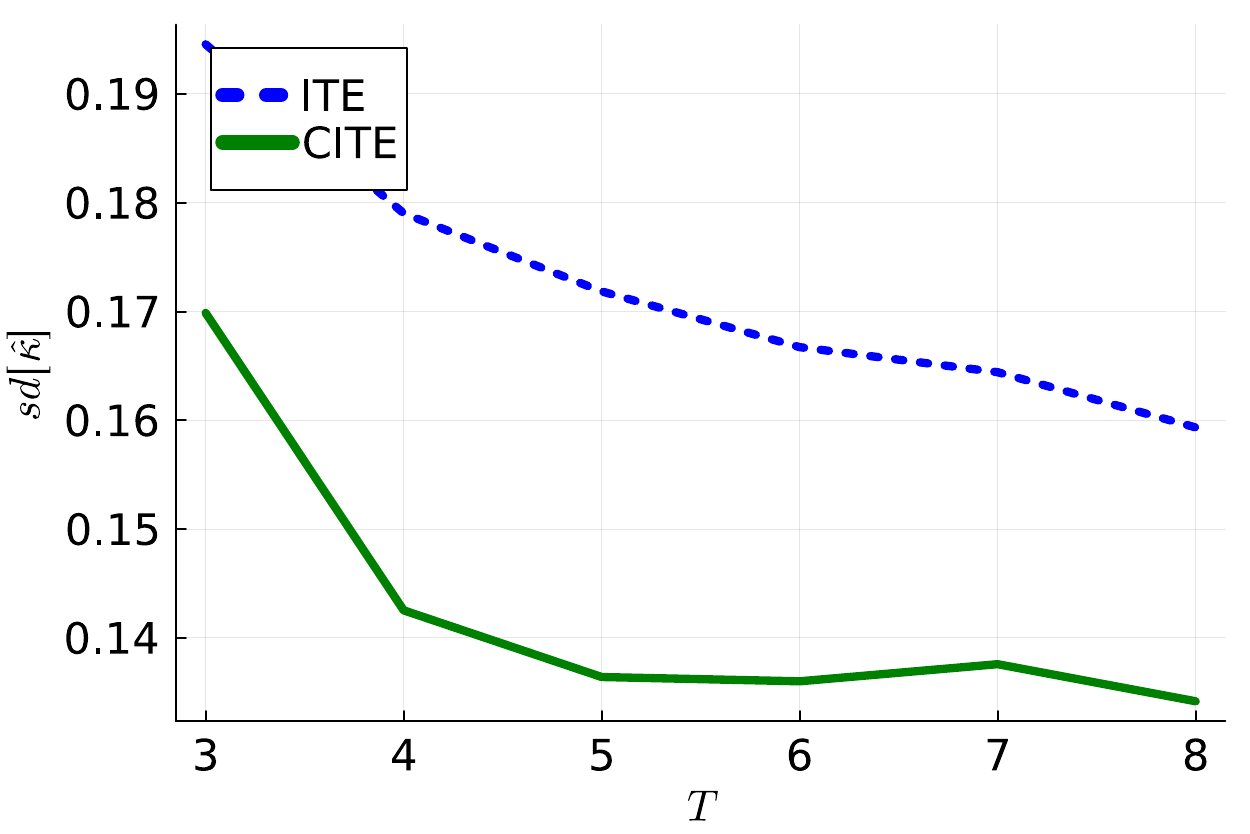}
        \caption{$\sigma_\epsilon = 2$}
        \label{fig:sd_e_2}
    \end{subfigure}
    \hfill
    \begin{subfigure}{0.48\textwidth}
        \centering
        \includegraphics[width=\textwidth]{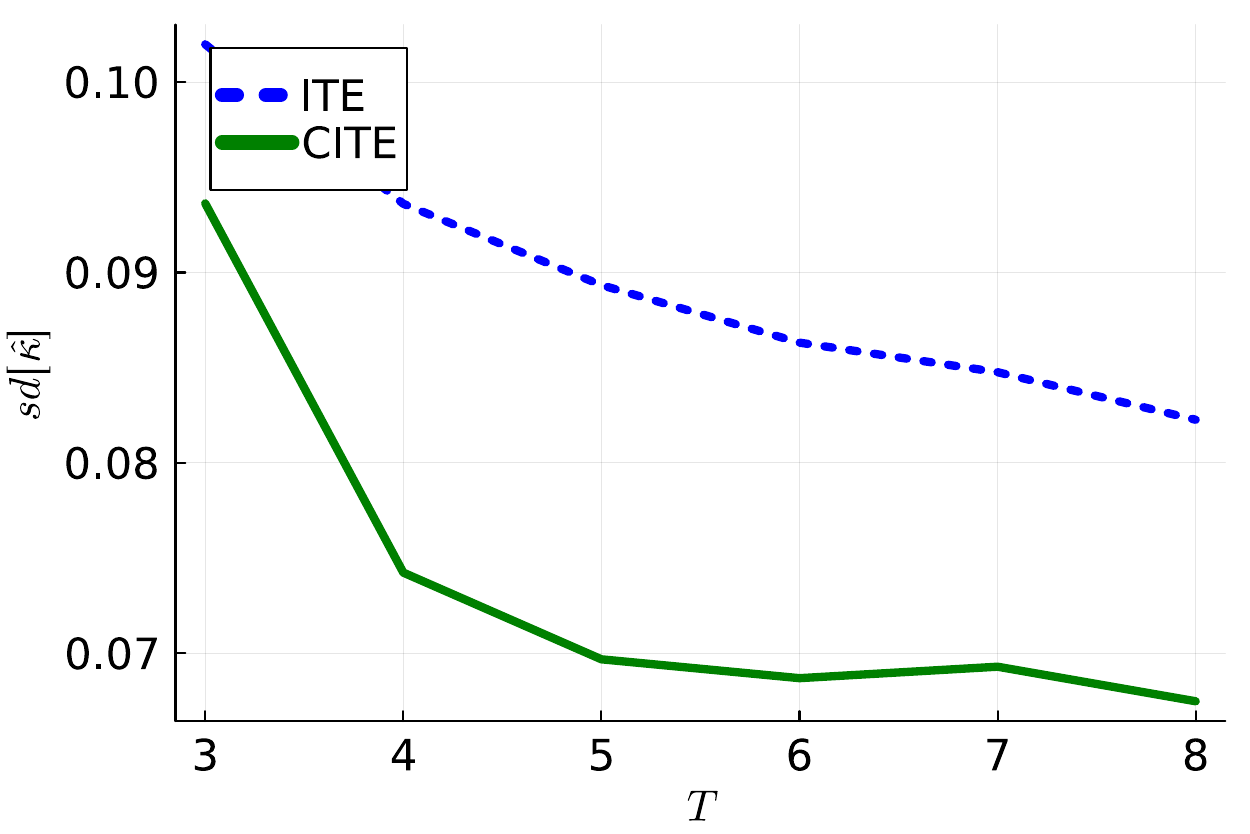}
        \caption{$\sigma_x = 2$}
        \label{fig:sd_x_2}
    \end{subfigure}

    \vspace{1em}  
    
    \begin{subfigure}{0.48\textwidth}
        \centering
        \includegraphics[width=\textwidth]{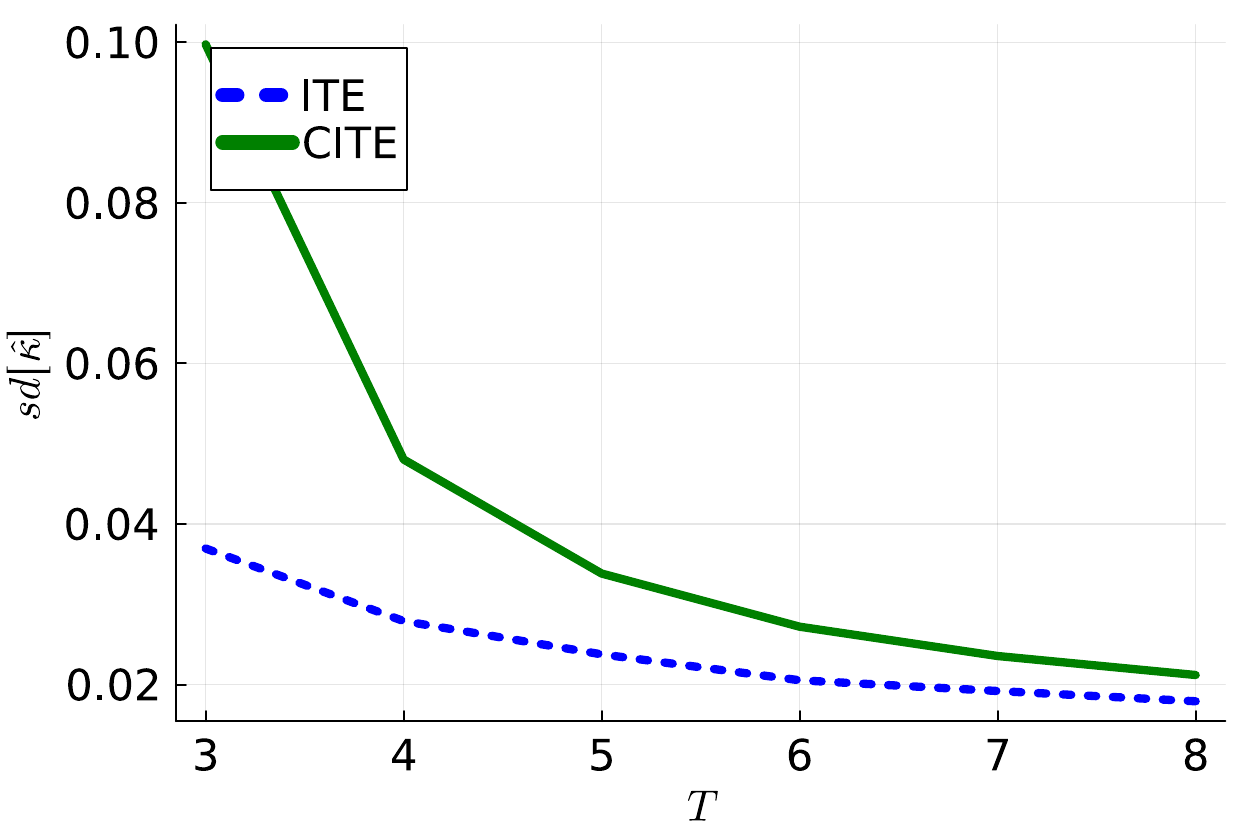}
        \caption{$\sigma_\epsilon = 0.1$}
        \label{fig:sd_e_01}
    \end{subfigure}
    \hfill
    \begin{subfigure}{0.48\textwidth}
        \centering
        \includegraphics[width=\textwidth]{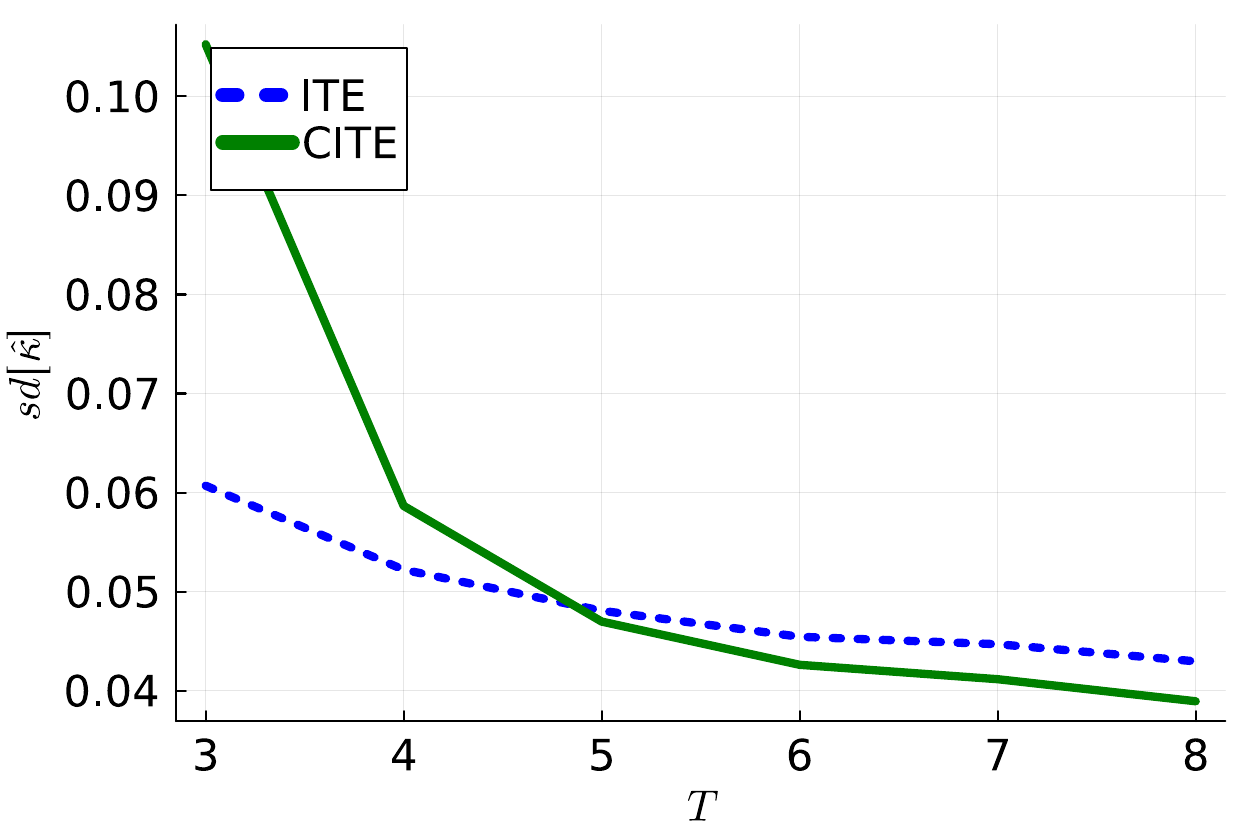}
        \caption{$\sigma_\epsilon = 0.5$}
        \label{fig:sd_e_05}
    \end{subfigure}
    \caption{The relative efficiency of CITE versus ITE changes depending on the design, and depends on $T$.}
    \label{fig:sd}
\end{figure}

Figure \ref{fig:sd_baseline} presents the baseline results. At $T=3$, the ITE is more efficient than CITE. At $T>3$, the CITE has a lower standard deviation. Figures \ref{fig:sd_u_2}--\ref{fig:sd_e_05} document that the profiles shift as we vary the values of $\sigma$. For example, an increase in effect heterogeneity $\sigma_\epsilon$ favors the CITE. For example, at $\sigma_\epsilon = 0.1$ (Figure \ref{fig:sd_e_01}), ITE is more efficient than CITE over the entire plotted range. Conversely, for $\sigma_\epsilon = 2$, CITE dominates ITE over $T$ (Figure \ref{fig:sd_e_2}).

Across all designs, the variability of both estimators decreases with $T$.
The relative efficiency of the two estimators depends on the design, 
and there is no clear ranking that holds across all designs.
Our main takeaway is that there is no clear penalty for using CITE to protect against bias.

\subsection{Inference}

In Section \ref{sec:variances}, we established simple inference procedures for ITE and CITE:
use cluster-robust standard errors for ITE, and use heteroskedasticity-robust standard errors in the second step of CITE.

\begin{figure}[htbp]
    \begin{subfigure}{0.48\textwidth}
        \centering
        \includegraphics[width=\textwidth]{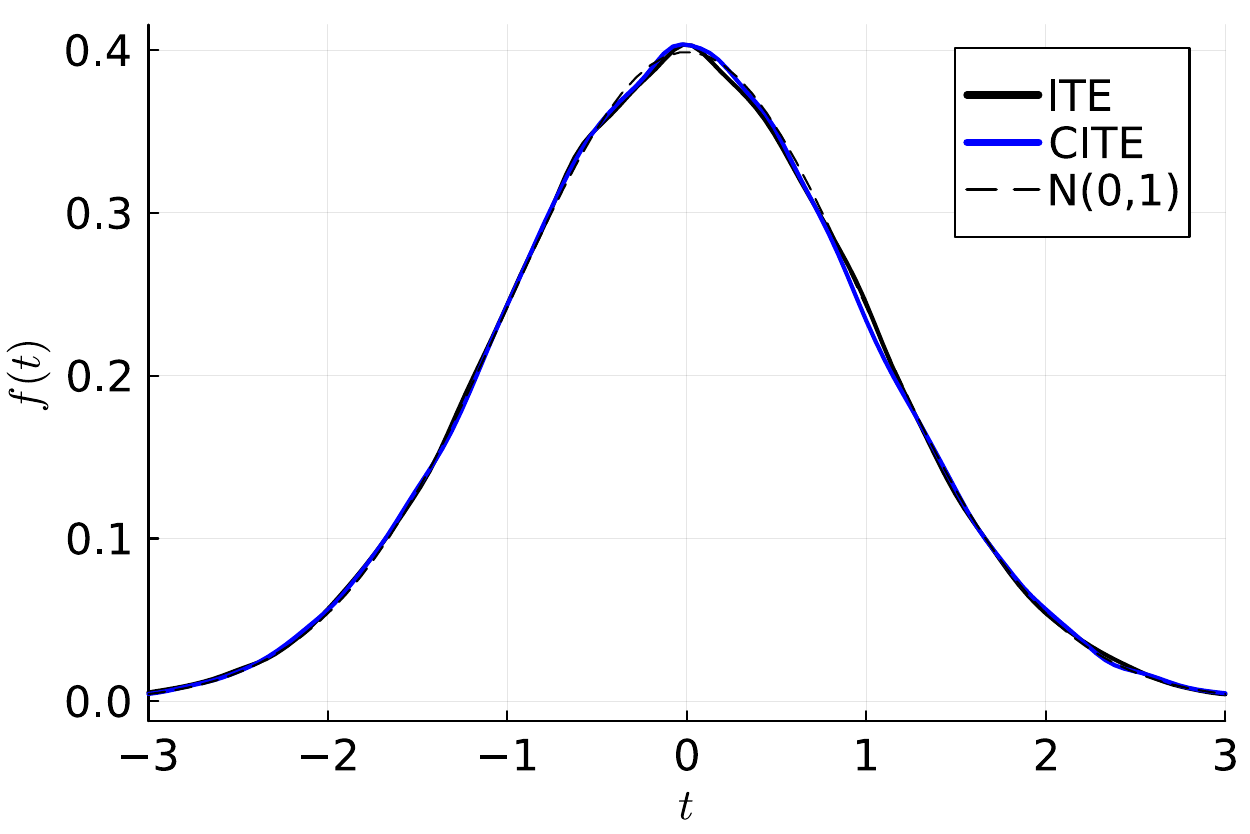}
        \caption{Distribution of $t$-values.}
        \label{fig:t}
    \end{subfigure}
    \hfill
    \begin{subfigure}{0.48\textwidth}
        \centering
        \includegraphics[width=\textwidth]{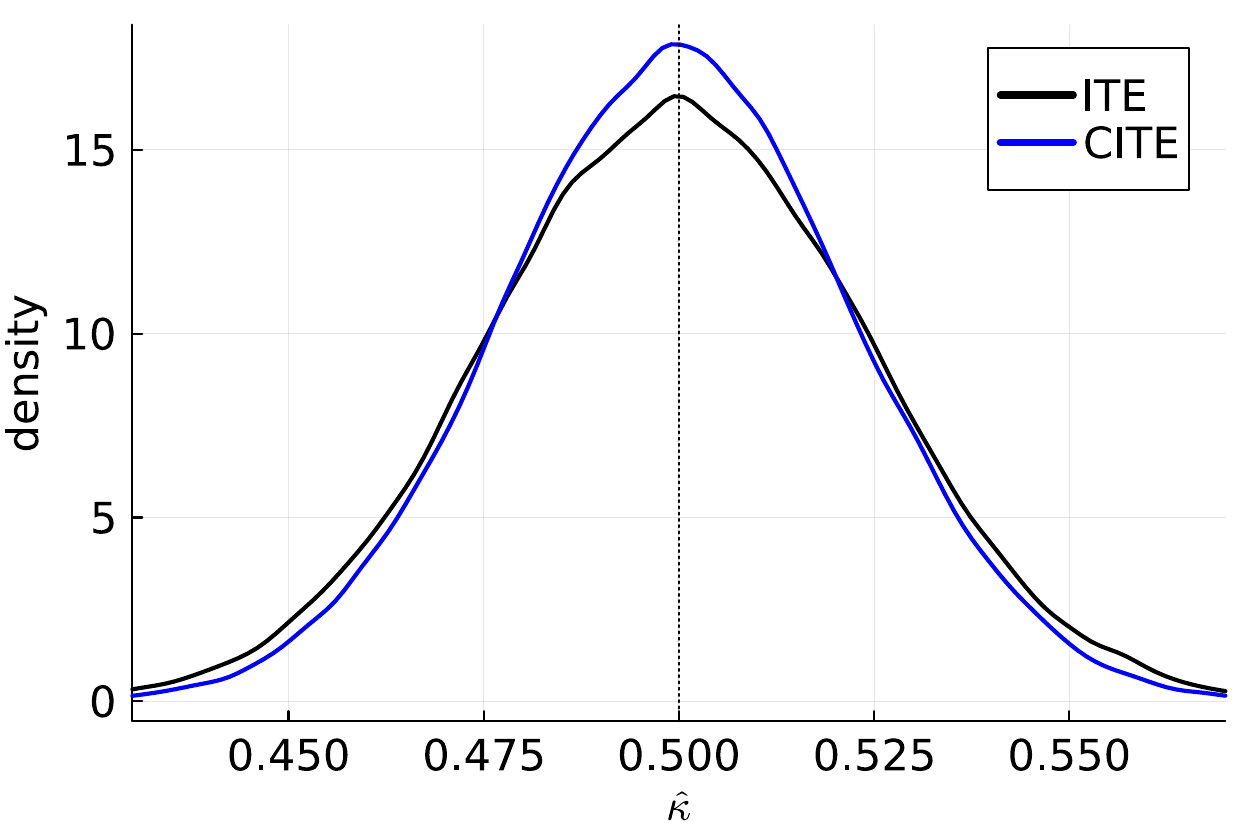}
        \caption{Distribution of ITE and CITE.}
        \label{fig:empirical_distribution}
    \end{subfigure}
    \caption{Validity of inference.}
    \label{fig:validity_inference}
\end{figure}

In both cases, the asymptotic distribution theory is standard, and standardized test statistics are asymptotically standard normal.
Figure \ref{fig:validity_inference} plots the density estimates of the standardized test statistics across 100000 simulations in a design with $n=1000$, $T=20$, $\kappa=0.5$, $\Delta=0$, and all $\sigma$ parameters equal to 1. 
The dashed line is the pdf of a standard normal distribution.
The dashed line is barely visible in the plot, because the distribution of $t-$statistics of both ITE and CITE are right on top. 
For the same design, Figure \ref{fig:empirical_distribution} plots the density estimates of ITE and CITE. 
They appear bell-shaped, with CITE less variable than ITE.
In conclusion, the results in Figure \ref{fig:validity_inference} are in line with the asymptotic theory.

\begin{figure}[htbp]
    \begin{subfigure}{0.48\textwidth}
        \centering
        \includegraphics[width=\textwidth]{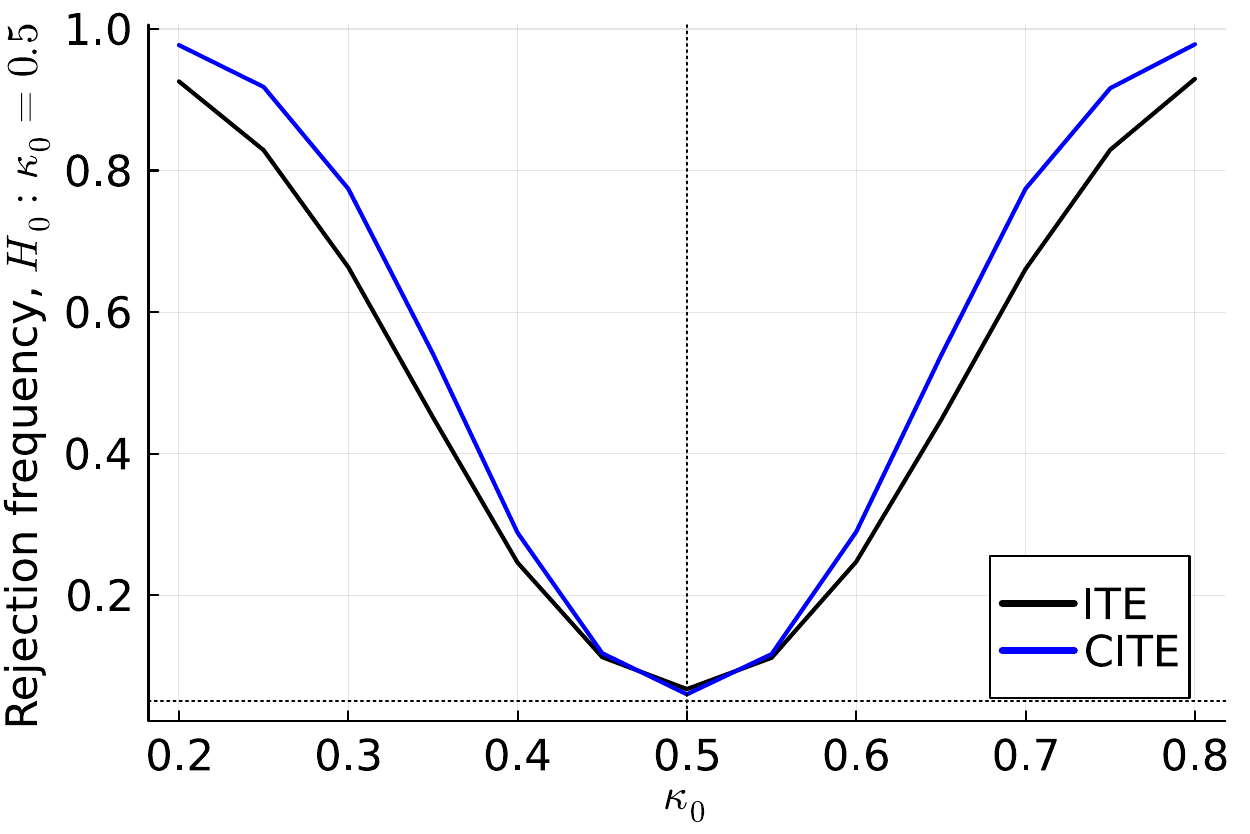}
        \caption{$n = 100$ and $\Delta = 0$.}
        \label{fig:rejection_frequency_baseline}
    \end{subfigure}
    \hfill
    \begin{subfigure}{0.48\textwidth}
        \centering
        \includegraphics[width=\textwidth]{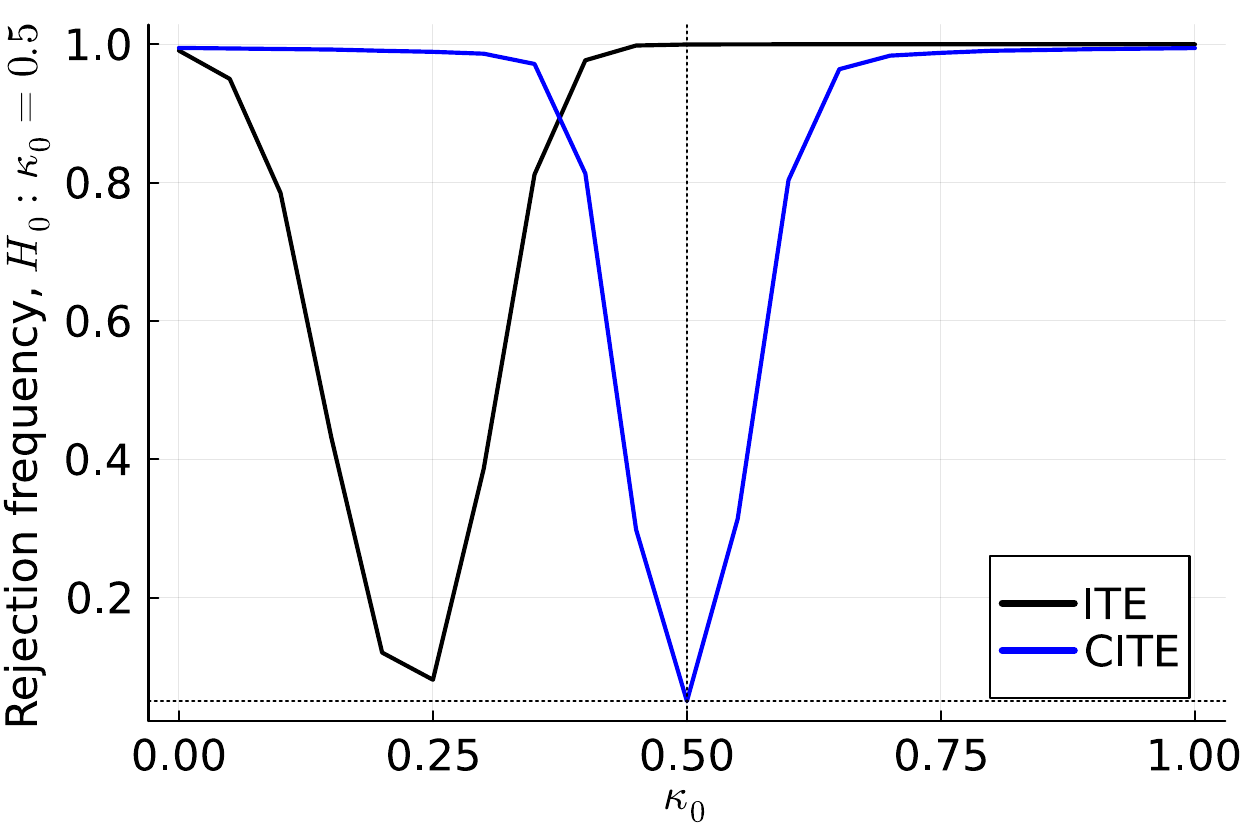}
        \caption{$n = 500$ and $\Delta = 0.3$.}
        \label{fig:rejection_frequency_ITE_inconsistent}
    \end{subfigure}
    \caption{Rejection frequencies for ITE and CITE.}
    \label{fig:rejection_frequencies}
\end{figure}

Figure \ref{fig:rejection_frequencies} provides further evidence on the asymptotic validity of our inferential procedure.
We plot the rejection frequencies for $H_0:\kappa_0 = 0.5$, varying $\kappa$ in the data generating process along the horizontal axis.
Both panels use 100000 simulations, and a nominal level of $\alpha = 0.05$  (indicated as a dashed horizontal line).
Figure \ref{fig:rejection_frequency_baseline} presents results for a design with $n=100$, $T=8$, $\Delta = 0$, and all $\sigma$ parameters equal to 1.
In this design, ITE is consistent. 
At $\kappa = 0.5$, rejection frequencies are close to the nominal level of $0.05$ for both estimators.
Rejection frequences away from $\kappa = 0.5$ are better for CITE: an incorrect null hypothesis is more frequently rejected.
Figure \ref{fig:rejection_frequency_ITE_inconsistent} presents results for a design with $n=500$, $T=8$, $\Delta = 0.3$, and all $\sigma$ parameters equal to 1.
In this design, the ITE curve is shifted left due to the bias in ITE under this design.
Compared to the left panel, due to the increased sample size, the CITE curve is steeper around 0.5, 
and the rejection frequency at $\kappa = 0.5$ is very close to $\alpha$.

\section{What does the relation between robots and employment depend on?} \label{sec:application}

The relationship between robots and employment has attracted considerable attention over recent years. \citet{Mondolo2022,  Juratetal2023}, and \citet{Restrepo2023} survey this literature, which is highly relevant beyond academia: a solid knowledge about the labor market consequences of technical change helps us understand whose jobs may be automated, and to design targeted policies. 

Yet, the empirical evidence on the robot-employment nexus is mixed to date. A meta-analysis by \citet{Guarascioetal2024} summarizes 33 studies with 644 estimates and finds considerable variation across studies and estimates, with a very small average negative relationship between robotization and employment. Within this literature, our application is related to cross-country industry level studies in the spirit of \citet{GraetzMichaels2018} and \citet{deVriesetal2020}.

One possible reason for the mixed findings about the robot-employment relationship is cross-country heterogeneity, as suggested by previous empirical results by \citet{Reljicetal2023} and \citet{ChenFrey2024}, for example. Such heterogeneity is highly plausible on theoretical grounds. Consider the seminal partial equilibrium model of \citet{AcemogluRestrepo2020jpe}, where employment $L$ in industry $t$ and region $i$ is affected by robotization $R$ through three channels: a negative labor displacement effect of robotization technology, a positive demand effect reflecting the productivity gains from robotization, and a composition effect, which captures that industries that robotize require more (non-automated) labor from non-expanding sectors. Formally, those three channels are captured in their equation: 
\begin{equation}
\label{eq:acemoglurestrepo2018jpe}
\frac{\Delta \operatorname{ln}L_{it}}{\Delta \operatorname{ln}R_{it}} = \underbrace{- \frac{\Delta \theta_t}{1- \theta_t}}_{displacement} + \underbrace{\frac{1}{\alpha} \Delta \operatorname{ln} Y_i}_{demand} - \underbrace{(\sigma + \frac{1}{\alpha} -1) \Delta \operatorname{ln}P^X_{it}}_{other},
\end{equation}

where $\theta_t$ is a technology parameter indicating the range of tasks that can be automated (i.e., performed by robots), $Y_c$ is output, $\sigma>0$ is the elasticity of substitution (between goods of different industries) and $P_{it}$ is the output price of industry $t$ in $i$, and $1-\alpha$ is the share of non-robot capital in the production process.\footnote{\citet{AcemogluRestrepo2020jpe} estimate a reduced-form version of equation \eqref{eq:acemoglurestrepo2018jpe} for US commuting zones $i$ but their theoretical model is more general and can be applied across countries $i$. Note that our subscript $t$ here indexes industries, not time periods. An earlier working paper version of our paper \citep{MurisWacker2022} contains a textbook application from \citep{stockIntroductionEconometrics2015} that illustrates how ITE and CITE can differ in a standard panel setting with $t$ indexing time periods.} Those effects may operate into opposing directions and cause marginal effect heterogeneity. Heterogeneity across industries is thoroughly documented by \citet{Bekhtiaretal2024}, while our application focuses on heterogeneity across countries.

We focus on two key candidates for effect heterogeneity across countries in the robot-employment relationship, which arise from the partial equilibrium model by \citet{AcemogluRestrepo2020jpe}: income p.c. levels and demand effects. The demand effect is straightforward and reflected in the second right-hand side term of equation \eqref{eq:acemoglurestrepo2018jpe}. The importance of the income p.c. level for the robot-employment relationship arises from the fact that \citet{AcemogluRestrepo2020jpe} assume the technology parameter $\theta$ to be homogeneous across $i$, which is plausible for US commuting zones but not across countries. Moreover, $\theta$ captures whether a task can \emph{technically} be automated. Whether this is \emph{economically} profitable depends on the cost savings from using robots ($\pi_i$ in \citealp{AcemogluRestrepo2020jpe}, who for simplicity assume $\pi_i>0 \ \forall i$). The cost-savings aspect unambiguously calls for a more negative effect of robotization on employment in higher-income countries: cost-saving potential $\pi_i$ positively depends on wage levels, which are higher in high-income countries.

The remainder of this section hence explores how income p.c. and its changes, reflecting demand effects, alter the robot-employment relationship. We will abstract from other factors that may give rise the cross-country effect heterogeneity in the robot-employment nexus for the sake of providing a traceable and instructive illustration of the performance of CITE and ITE in a relevant applications with plausible interaction term effects. Code documentation is available on the authors' GitHub repository (https://github.com/KMWacker/CITErobots).\\

\subsubsection*{Regression setup: ITE vs. CITE}

In cross-country industry-level studies, the relationship between labor market outcomes $L$ and robots is typically estimated as a long-run first-difference equation (e.g., \citealp{GraetzMichaels2018, deVriesetal2020, Bekhtiaretal2024}):

\begin{equation}
\label{eq:robotindustry}
\Delta \operatorname{ln} L_{it} = \beta \Delta Robots_{it} + a_i + \varepsilon_{it}.
\end{equation}

The inclusion of country-specific fixed effects $a_i$ is convenient for isolating country-specific employment trends that are associated with robots. 
To capture the effect heterogeneity suggested above, we interact $\Delta Robots$ with the relevant interaction terms $H_1$ (income levels) and $H_2$ (demand changes) in an ITE settting:

\begin{equation}
\label{eq:robotindustry_ite}
\Delta \operatorname{ln} L_{it} = \beta \Delta Robots_{it} + \sum_{k=1}^K \kappa_{k} \Delta Robots_{it} \cdot H_{k,i} + a_i + \nu_{it},
\end{equation}
\noindent where our (up to) $K=2$ interaction terms are country-specific income p.c. levels and demand changes. Recall that $\beta$ in this framework corresponds to $\kappa_0$ in the context of CITE.

The standard ITE approach consists in estimating equation \eqref{eq:robotindustry_ite} through least squares (augmented with dummy variables $\alpha_i$). As thoroughly discussed in Section \ref{sec:additional_results}, this requires strong exogeneity assumptions for consistent estimation of the interaction term coefficients $\kappa_1, \kappa_2$. In particular, we may be concerned about two cases that give rise to a correlation between the error term $u$ and the regressors.

One reason for concern is the behaviour of estimators in the presence of omitted interaction variables. Changes in demand $H_{2i}$ are likely correlated with income levels $H_{1i}$ and both may give rise to effect heterogeneity, leading to a bias in $\hat{\kappa}$ if one of them is omitted. We will hence assess to what extent it matters for ITE and CITE if $\kappa_1 H_1$ and $\kappa_2 H_2$ are jointly or individually included to gauge how both estimators behave in the presence of potentially omitted interaction variables.

A second reason for concern is more fundamental and concerns unobservable effect heterogeneity. For example, robots are frequently used for manual routine work in many industries, like health care. Patients, or clients more generally, in more technology-minded countries may be more open to being supported by robots. This gives rise to a higher worker substitutability, implying that the relationship between employment and robots is more negative in tech-minded countries. Tech-mindedness, which is not directly observable, will hence enter the error term $u$ in our ITE regression equation \eqref{eq:robotindustry_ite} (through $X_{it}\epsilon_i$ in our general notation, see eq. \eqref{eq:outcome-equation}-\eqref{eq:heterogeneity-equation}). Since tech-mindedness plausibly gives rise to higher robotization in the first place, the error term $u$ and $\Delta Robots$ will be positively correlated in equation \eqref{eq:robotindustry_ite}, which is a clear violation of the ITE exogeneity assumption (see Section \ref{sec:correct_ITE_fail}).

Conversely, CITE explicitly models such country-specific effects. In the context of our application, CITE consists in first estimating the unit-specific parameters $\beta_i$ for each individual country $i$ through the outcome equation:

\begin{equation}
\label{eq:robotindustry_cite}
\Delta \operatorname{ln} L_{it} = \beta_{i} \Delta Robots_{it} + a_i + u_{it},
\end{equation}
\noindent and subsequently exploring their relationship with country-specific income p.c. levels $H_1$ and demand changes $H_2$ by running the heterogeneity equation:

\begin{equation}
\widehat{\beta}_i = \kappa_0 + \kappa_1 H_{1i} + \kappa_2 H_{2i} + \epsilon_i.
\end{equation}

Both approaches, ITE and CITE, allow us to investigate if previous empirical studies have omitted an essential factor how robots influence labor market outcomes and facilitate a comparison of the performances of ITE and CITE.

It is important to recall that CITE absorbs all country-specific effect heterogeneity in the robot-employment relation in $\beta_{i}$, while ITE only allows for heterogeneity in the modeled $interation$ terms. Furthermore, note that equations \eqref{eq:robotindustry} - \eqref{eq:robotindustry_cite} are long-run first-difference equations, which means that industry fixed effects are ``differenced away'' and that our subscript $t$ now indexes industries, not time periods, to illustrate the applicability of CITE in various panel setups. A standard textbook panel application (as well as a higher-dimensional panel application) can be found in an earlier working paper version of our paper \citep{MurisWacker2022}.

\subsubsection*{Data}

We use data from 15 manufacturing industries across 35 countries that is explained in full detail in \citet{DijkstraWacker2025}. Robot stocks are calculated from IFR data (International Federabtion of Robotics) with a perpetual inventory method and divided by thousands of employees in the respective industry (from OECD TiM, see below). As suggested by \citet{GraetzMichaels2018}, using raw or log changes in this robot density is not recommendable because of very low (or zero) starting values in the mid-2000s. We hence follow their recommendation to construct percentiles of (employment-weighted) changes in the robot stock.

For employment numbers, we rely on the 2023 edition of OECD's Trade in Employment (TiM) database (variable EMPN). Additionally, we merge the country-specific log of gross domestic product per capita, ln GDP p.c. (rgdpo/emp), from the Penn World Tables 10.0 \citep{feenstrapwt9}, which captures income levels across countries and, in first differences, demand changes. Those are our interaction variables $H_1, H_2$, as suggested by the first two right-hand side terms of equation \eqref{eq:acemoglurestrepo2018jpe}.

For $\Delta \operatorname{ln} L$ and $\Delta Robots$, we construct 10-year changes comparing the 2014-2018 averages to 2004-2008 averages.\footnote{Comparing 10-year changes is standard in this literature. We construct 5-year averages at the start- and end-point to avoid year-specific variation in key variables (e.g., due to the global financial crisis in 2008).} For ln GDP p.c. we take country-specific averages over the 2004-2018 period. For changes in demand, we construct changes in ln GDP p.c. between the 2014-2018 and 2004-2008 averages (divided by 10 to obtain annual approximations). Further details on the data construction can be inferred from the code in the associated GitHub repository `CITErobots'.

\begin{table}[h!tbp]\centering
\def\sym#1{\ifmmode^{#1}\else\(^{#1}\)\fi}
\caption{Summary Statistics} \label{tab:robot_sumstat}
\begin{tabular}{l*{1}{ccccc}}
\hline\hline
                    &           N&        mean&          sd&         min&         max\\
\hline
$\Delta$ ln L           &            &      -0.113&       0.288&       -1.39&        0.92\\
$\Delta$ Robots &            &       0.500&       0.290&        0.00&        1.00\\
ln GDP p.c.    &            &      11.165&       0.438&        9.77&       11.94\\
$\Delta$ demand     &            &       0.018&       0.016&       -0.01&        0.06\\
\hline
Observations        &         509&            &            &            &            \\
\hline\hline
\end{tabular}
\end{table}

\begin{figure}[ht]
\caption{Changes in employment and robotization}
\label{fig:descriptivescatter}
\includegraphics[scale=0.78]{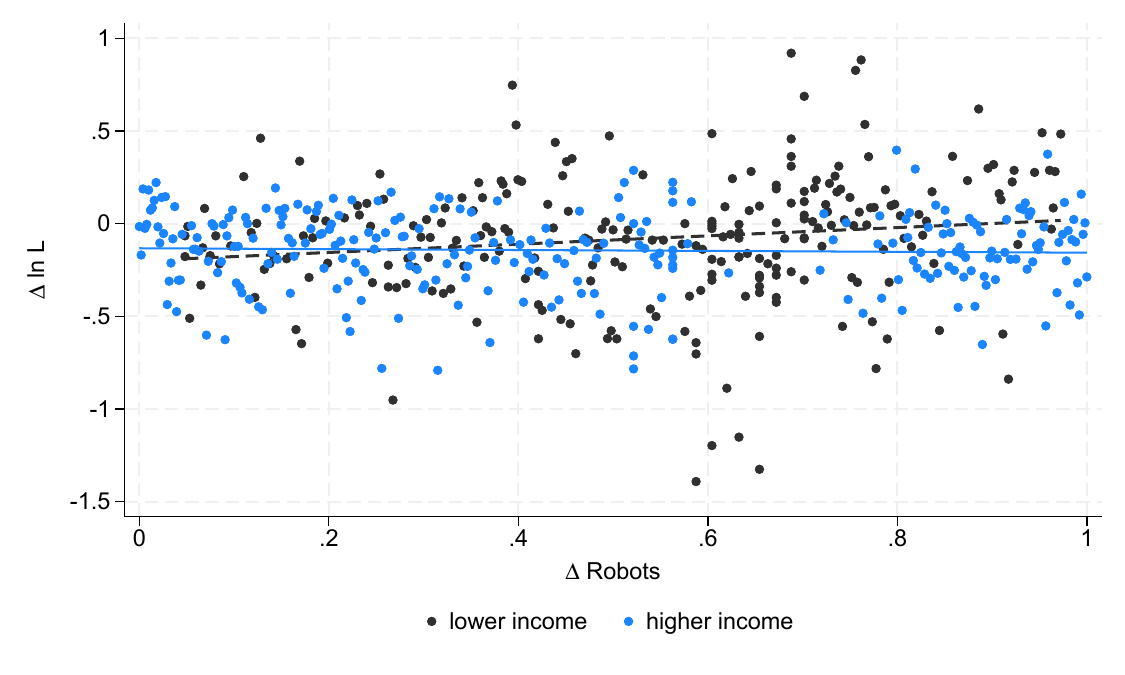} 
\end{figure}

Table \ref{tab:robot_sumstat} reports the summary statistics of our sample, while Figure \ref{fig:descriptivescatter} plots our key variables $\Delta \operatorname{ln} L$ and $\Delta Robots$ in a scatter that differentiates between lower- and higher-income countries. A number of features is worth highlighting. First, employment declines, on average (unweighted), with considerable heterogeneity: the standard deviation is more than double the mean. Second, since $\Delta Robots$ is constructed as percentile change, it ranges from 0 to 1 by construction, with a mean and IQR of 0.5. GDP p.c. is rather high, reflecting that most of our sample countries are high-income. Its rather low standard deviation reflects that this variable does not vary in the $t$ dimension of our panel (i.e., across industries of the same country). Figure \ref{fig:descriptivescatter} separately illustrates countries at or above the ln GDP p.c. average (blue color) and those below average ln GDP p.c. (black). Third, there is some negative correlation between $\Delta Robots$ and ln GDP p.c., although this is not obviously visible from Figure \ref{fig:descriptivescatter}. Their correlation coefficient is -0.10 (not reported) and averages of $\Delta Robots$ are slightly higher for lower-income countries than for higher-income countries (0.54 vs. 0.47). Fourth, Figure \ref{fig:descriptivescatter} suggests a positive descriptive correlation between $\Delta \operatorname{ln} L$ and $\Delta Robots$ in lower-income countries (dashed black line) and no correlation in higher-income countries (solid blue line), which our estimation will explore more thoroughly. Fifth, the sample is slightly unbalanced (with 509 observations), reflecting missing values in 1-4 industries across 9 sample countries. Finally, it is worth noticing that changes in demand are close to the historic average GDP growth rate of 2\% per annum documented in \citet{PritchettSummers2014} and negatively correlated with ln GDP p.c. (-0.64, not reported), consistent with the idea of unconditional convergence \citep{pateletal2021}. 

Note that our third and fourth observation, taken together and at face value, may constitute a problem for ITE if an interaction of $\Delta Robots$ with ln GDP p.c. is omitted: $X$ ($\Delta Robots$) is plausibly correlated with the error term $\epsilon$ in the heterogeneity equation \eqref{eq:heterogeneity-equation}. This violates the required Assumption \ref{assu:Exogeneity-1-strengthened} for consistency of ITE and is likely on economic grounds: richer countries may exhibit certain features that simultaneously affect robotization and the robot-employment nexus.

\subsubsection*{Results}

Table \ref{tab:robot_regressions} summarizes our regression results, with ITE in the upper panel A and CITE in the lower panel B. We start with a simple linear regression of $\Delta \operatorname{ln}L$ on $\Delta Robots$ as a baseline in column (1), which facilitates comparison to the literature. In our (manufacturing) sample, there seems to be an overall positive relationship between industries' robotization and employment changes.

In column (2) of Table \ref{tab:robot_regressions}, we allow this relationship to vary with the interaction term for income levels (ln GDP p.c., $H_1$). In line with the above theoretical considerations (and already suggested by Figure \ref{fig:descriptivescatter}), we find a negative interaction term coefficient $\hat{\kappa}_1$: the higher a country's income p.c. (and hence wage) level, the less favorable employment outcomes are associated with robotization. We will revert to the estimated marginal effects (which are graphically depicted in Figure \ref{fig:robotscatter}) below but highlight a higher (absoulte) interaction term coefficient estimate for CITE, compared to ITE. Both estimators are relatively precise (interaction term standard errors $<1.7 \times |\hat{\kappa}_1|$), especially ITE.

\begin{table}
\centering
{
\caption{Regression results} \label{tab:robot_regressions}
\def\sym#1{\ifmmode^{#1}\else\(^{#1}\)\fi}
\begin{tabular}{l*{4}{c}}
\hline\hline
            &\multicolumn{1}{c}{(1)}&\multicolumn{1}{c}{(2)}&\multicolumn{1}{c}{(3)}&\multicolumn{1}{c}{(4)}\\
            &\multicolumn{1}{c}{$\Delta$ ln L}&\multicolumn{1}{c}{$\Delta$ ln L}&\multicolumn{1}{c}{$\Delta$ ln L}&\multicolumn{1}{c}{$\Delta$ ln L}\\
            \hline

\multicolumn{5}{c}{}\\
\multicolumn{5}{l}{\emph{Panel A: ITE results}}\\
\multicolumn{5}{c}{}\\
$\Delta$ Robots &       0.125***&       5.267***&      -0.008   &       4.053** \\
            &      (0.05)   &      (1.73)   &      (0.06)   &      (1.97)   \\
$\Delta$ Robots $\times$ ln GDP p.c. &               &      -0.454***&               &      -0.352** \\
            &               &      (0.15)   &               &      (0.17)   \\
$\Delta$ Robots $\times \ \Delta$ demand &               &               &      10.475***&       4.527   \\
            &               &               &      (3.77)   &      (5.03)   \\
\hline
Country FEs & yes & yes & yes & yes \\
r2          &       0.308   &       0.326   &       0.320   &       0.327   \\
N           &     509   &     509   &     509   &     509   \\

\hline\hline
\multicolumn{5}{c}{}\\
\multicolumn{5}{l}{\emph{Panel B: CITE results}}\\
\multicolumn{5}{c}{}\\
$\Delta$ Robots      &       0.449** &      13.098*  &       0.214   &      16.380*  \\
            &      (0.18)   &      (7.33)   &      (0.16)   &      (9.36)   \\
$\Delta$ Robots $\times$ ln GDP p.c.   &               &      -1.133*  &               &      -1.408*  \\
            &               &      (0.65)   &               &      (0.81)   \\
$\Delta$ Robots $\times \ \Delta$ demand    &               &               &      13.271   &     -11.892   \\
            &               &               &      (9.60)   &     (14.99)   \\
\hline
Country FEs & yes & yes & yes & yes \\
r2 (second stage)         &       N/A  &       0.209   &       0.037   &       0.226   \\
N           &      35   &      35   &      35   &      35 \\
\hline\hline
\multicolumn{5}{l}{\footnotesize * p$<$0.1, ** p$<$0.05, *** p$<$0.01}\\
\end{tabular}
}
\end{table}

In column (3) of Table \ref{tab:robot_regressions}, we explore the hypthesis that heterogeneities in the robot-employment nexus are driven by demand changes $H_2$. Consistent with theory both point estimates of the interaction term coefficient suggest a more positive association between robotization and employment growth the more aggregate demand increases in a country. However, ITE would leave the researcher highly confident in this specification that the partial effect between robots and employment linearly depends on demand effects. The associated ITE p-value $< 0.01$ in column (3) suggests that the data are quite compatible with the specified interaction model, while CITE inference raises considerably more caution. Also notice the low R squared of the second stage CITE projection in column (3).

Column (4), which includes both interaction terms, illustrates that the demand channel identified in column (3) is most likely due to an omitted variable bias. Recall that demand changes and income p.c. levels are inversely correlated across countries (-0.64): richer countries experience lower increases in demand. The interaction term $\Delta Robots \times \Delta demand$ in column (3) hence captures all cross-country heterogeniety that is correlated with demand changes, including substantial heterogeneity due to income p.c. levels. It is, of course, highly likely that $\Delta demand$ (and ln GDP p.c.) are correlated with other sources of effect heterogeneity in the robot-employment nexus.\footnote{Likewise, it is plausible that our measure for demand changes is erroneous -- which substantiates our point that ITE inference in column (3) of panel A in Table \ref{tab:robot_regressions} is problematic.} As we know from our Monte Carlo inference simulations, ITE's rejection probabilities for t-tests of a null hypothesis about the interaction term coefficient can be highly distorted in this case. The stark ITE-related differences in p-values for such rejections when moving from column (3) to column (4) in panel A of Table \ref{tab:robot_regressions} should be a warning signal for applied researchers that inference of standard ITE may be highly susceptible and misleading.

\subsection*{Marginal effects}

\begin{figure}
\caption{Marginal effects of robots on employment}
\label{fig:robotscatter}
\includegraphics[scale=0.78]{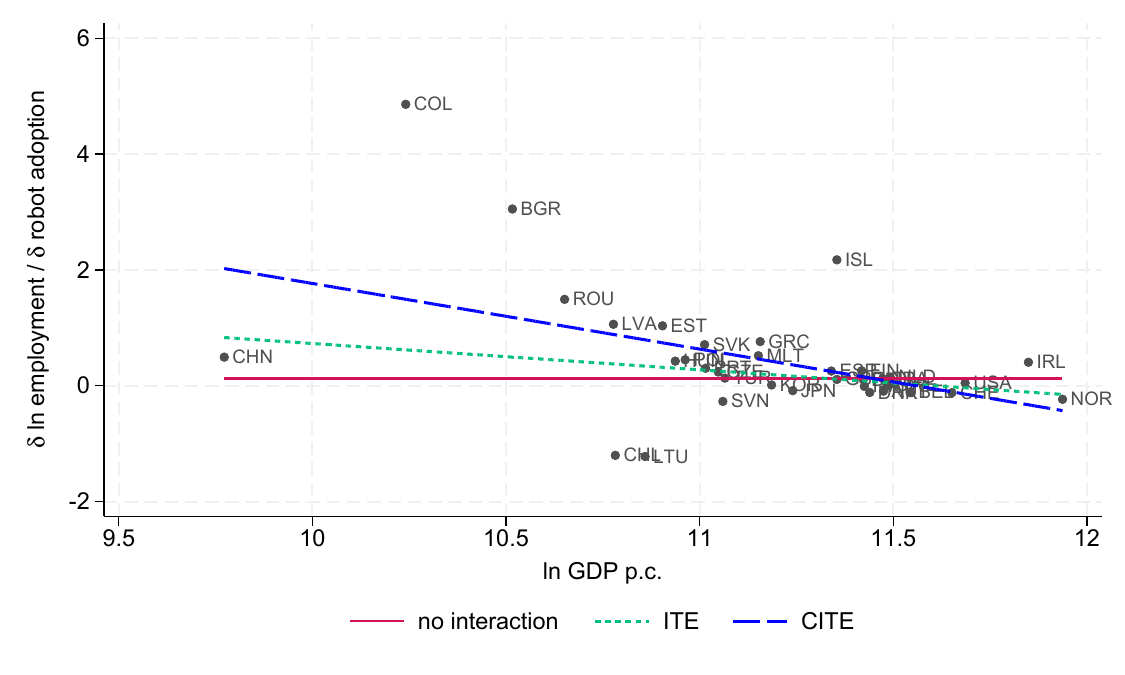} 
\end{figure}

The results in Table \ref{tab:robot_regressions} suggest that cross-country heterogeneity in the robot-employment relationship is to a significant degree driven by differences in income per capita (and associated wage) levels. Column (2) of panel B suggests that 21\% of cross-country heterogeneity can be explained by variation of ln GDP p.c. (see also Figure \ref{fig:robotscatter}).

How much of a difference does the income level make for employment effects of robotization?\footnote{In this section, we treat the estimated relationship as an `effect', well-aware that other sources of endogeneity may pose challenges to a causal interpretation.} And how would that estimated marginal effect vary between ITE and CITE? Figure \ref{fig:robotscatter} is instructive for answering this question. Let us consider income levels of ln GDP p.c. = (10.5, 11.4, 11.7). In our sample, the former corresponds to Bulgaria, which is at the lower end of new European Union member states and a plausible level for various middle-upper income countries (e.g., between Colombia and Chile). The second corresponds to Germany, the latter to the United States. Recall that an overall regression with no interaction effects would suggest a positive relationship equal to 0.13 in either case (red line in Figure \ref{fig:robotscatter} and column (1) in panel A of Table \ref{tab:robot_regressions}).

Considering a difference in robotization rates of $\Delta Robots = 0.5$ and estimates from column (2) of Table \ref{tab:robot_regressions}, ITE suggests an associated employment increase of 25\% over a decade, for a country at a ln GDP level of 10.5.\footnote{Recall that $\Delta Robots$ is constructed as percentile changes, ranging from 0 to 1. The IQR of $\Delta Robots = 0.5$ hence compares the 25th percentile of robot adoption to the 75th percentile of robot adoption. Estimated marginal effect: $0.5 \cdot (\hat{\kappa}_0 + \hat{\kappa}_1 \operatorname{ln} GDP p.c.)$.} The magnitude is much larger with CITE, 60\%, and Figure \ref{fig:robotscatter} suggests that it more plausibly captures the estimated country-specific coefficients for various upper-middle income countries, while ITE seems heavily impacted by the low country-specific estimate of China, the lowest-income country in our sample.

For a ln GDP p.c. level of 11.4 (approximately Germany), the estimated marginal effects of CITE (9.1\%) and ITE (4.6\%) are both positive and relatively close to the magnitude without interaction effect (6\% $ = 0.5 \cdot 0.13$, cf. column (1) panel A of Table \ref{tab:robot_regressions}). For a ln GDP p.c. level of 11.7 (approximately US), the estimated marginal effects are negative, and stronger for CITE (-7.9\%) than for ITE (-2.2\%). Considering that CITE is consistent under milder assumptions than ITE, our results suggest that ITE and homogeneous estimation without interaction terms both underestimate the negative effects of robotization for manufacturing employment in countries at the highest income levels in the sample (ln GDP p.c. $>$ 11.57; Ireland, Norway, Switzerland, United States).

If income (and associated wage) levels are indeed a main source of robot-employment effect heterogeneity, this may reconcile conflicting country level evidence in the literature. For example, \citet{AcemogluRestrepo2020jpe}, find
negative consequences of robot exposure on regional labor markets in the US, while \citet{Dauthetal2021} find no such aggregate employment effects for Germany, despite using a similar approach. This is quite consistent with our quantitative estimates about effect heterogeneity due to income differences across countries (especially when considering differences in sample coverage and methodologies).

CITE, of course, provides country-specific estimates of the robot-employment relationship, as they are illustrated in Figure \ref{fig:robotscatter}. The reason why we do not consider those country-specific estimates for our marginal effect calculations is that they suffer from incidental parameter bias. The purpose of our paper is to estimate interaction terms in exactly such a short-$T$ panel, where country-specific estimation is no option, and we have shown that CITE offers a consistent projection of those (individually biased) panel unit estimates on the interaction variable of interest. We hence consider marginal effects of a random country at an income level like, e.g., Bulgaria, rather than a Bulgaria-specific coefficient. Applied analysts working on specific countries may be willing to accept the incidental parameter bias in country-specific estimates, but this is not the purpose of our paper.

\subsection*{Application takeaways}

An important insight of our analysis for the literature on employment effects of robotization is that economic displacement effects of robots are plausibly stronger in countries at higher income levels, where wages are usually higher. Econometrically, our application suggests that this income-interaction effect is underestimated by standard ITE and that ITE may lead to premature rejection of a null hypothesis about an interaction term in the presence of misspecification (e.g., when another relevant interaction term is omitted).

Our application results, and particularly Figure \ref{fig:robotscatter}, also highlight why  CITE is attractive for applied econometricians. Because unit-specific coefficients are estimated in a first step and then further analyzed in a second step (through projection on the relevant interaction variables $H$), this offers a lot of flexibility and diagnostics tools in the second step.\footnote{For example, one may speculate to what extent Bulgaria, Colombia, China, and Chile are outliers. Indeed, those countries surpass a critical value for `Cook's distance' and when they are excluded from the analysis, we obtain ITE and CITE interaction coefficients for ln GDP p.c. of -0.58 and -0.59 (significant at the 1 and 10\% level), respectively, which are very close to each other and fall between the point estimates reported in column (2) of Table \ref{fig:robotscatter}.}

\section{Conclusion}\label{sec:conclusion}

Our paper has highlighted that the most common approach to estimate interaction effects with panel data neglects residual heterogeneity in the effect of $X$ on $Y$ across panel units. If this unobserved effect heterogeneity is systematically correlated with explanatory variables in the model, which is plausible in most econometric applications, this standard approach will not provide consistent estimates of the interaction effects. 

We advocate for explicit modeling of effect heterogeneity across panel units with a correlated interaction term estimator (CITE) that requires less demanding exogeneity assumptions for consistency. Those findings are irrespective of whether interaction variables are constant within panels or not.

Our argument in favor of CITE resembles the case for additive fixed effects in linear panel data models. While additive fixed effects explicitly capture unobserved heterogeneity in unit-specific levels of $Y$, CITE explicitly models the effect heterogeneity $\beta_i$ that most common panel interaction estimates do not account for. 

Our paper has focused on the comparison of ITE and CITE for the estimation of interaction effects in panel data. Two alternative approaches that are sometimes used in empirical practice are (i) subgroup analysis -- in which one conducts the analysis conditional on every value of the interaction variable (e.g. \cite{chettyEffectsExposureBetter2016a}) -- and (ii) saturating the heterogeneity model and then running ITE (e.g., \cite{norrisEffectsParentalSibling2021}). Either approach may bridge the gap between ITE and CITE. Although such approaches may work well when the interaction variables are discrete, they are not feasible or practical for continuous interaction variables. It would be interesting to extend our analysis to include a comparison with these alternative approaches.

An obvious extension of our research is to derive a test statistic for consistency of ITE when compared to CITE. Until such a test becomes available, we strongly recommend the use of CITE or at least reporting CITE results next to ITE results. Given the simple implementation of CITE, this should become standard in applied econometrics when estimating interaction terms in panel data.\footnote{If one is interested in interactions with varying variables $G$, it is sufficient to additionally include an interaction of the $X$ variable(s) with panel dummy variables. If one is interested in interactions with $H$ variables that are constant within panels, those panel specific estimates from the first step need to be saved and regressed on the $H$ variables in a second step. An adjustment of standard errors is necessary to account for first-step estimation.}

An additional advantage of CITE, beyond its less restrictive exogeneity assumptions, is that the unit-specific effect heterogeneity captured in $\beta_i$ can provide further insights about sources of heterogeneity in panel data. Recent advances in machine learning appear particularly promising to further explore those first-step CITE estimates of unit-specific effects.

\newpage
\bibliographystyle{econometrica.bst}
\bibliography{Konstantin}

\begin{thebibliography}{}

\bibitem [\protect \citeauthoryear {%
Acemoglu%
\ \BBA {} Restrepo%
}{%
Acemoglu%
\ \BBA {} Restrepo%
}{%
{\protect \APACyear {2020}}%
}]{%
AcemogluRestrepo2020jpe}
\APACinsertmetastar {%
AcemogluRestrepo2020jpe}%
\begin{APACrefauthors}%
Acemoglu, D.%
\BCBT {}\ \BBA {} Restrepo, P.%
\end{APACrefauthors}%
\unskip\
\newblock
\APACrefYearMonthDay{2020}{}{}.
\newblock
{\BBOQ}\APACrefatitle {Robots and Jobs: Evidence from US Labor Markets} {Robots
  and jobs: Evidence from us labor markets}.{\BBCQ}
\newblock
\APACjournalVolNumPages{Journal of Political Economy}{128}{6}{2188-2244}.
\newblock
\begin{APACrefDOI} \doi{10.1086/705716} \end{APACrefDOI}
\PrintBackRefs{\CurrentBib}

\bibitem [\protect \citeauthoryear {%
Alfaro%
, Chanda%
, {Kalemli-Ozcan}%
\BCBL {}\ \BBA {} Sayek%
}{%
Alfaro%
\ \protect \BOthers {.}}{%
{\protect \APACyear {2004}}%
}]{%
alfaroFDIEconomicGrowth2004}
\APACinsertmetastar {%
alfaroFDIEconomicGrowth2004}%
\begin{APACrefauthors}%
Alfaro, L.%
, Chanda, A.%
, {Kalemli-Ozcan}, S.%
\BCBL {}\ \BBA {} Sayek, S.%
\end{APACrefauthors}%
\unskip\
\newblock
\APACrefYearMonthDay{2004}{}{}.
\newblock
{\BBOQ}\APACrefatitle {{{FDI}} and Economic Growth: The Role of Local Financial
  Markets} {{{FDI}} and economic growth: The role of local financial
  markets}.{\BBCQ}
\newblock

\PrintBackRefs{\CurrentBib}

\bibitem [\protect \citeauthoryear {%
Alfaro%
, Chanda%
, {Kalemli-Ozcan}%
\BCBL {}\ \BBA {} Sayek%
}{%
Alfaro%
\ \protect \BOthers {.}}{%
{\protect \APACyear {2010}}%
}]{%
alfaroDoesForeignDirect2010}
\APACinsertmetastar {%
alfaroDoesForeignDirect2010}%
\begin{APACrefauthors}%
Alfaro, L.%
, Chanda, A.%
, {Kalemli-Ozcan}, S.%
\BCBL {}\ \BBA {} Sayek, S.%
\end{APACrefauthors}%
\unskip\
\newblock
\APACrefYearMonthDay{2010}{}{}.
\newblock
{\BBOQ}\APACrefatitle {Does Foreign Direct Investment Promote Growth?
  {{Exploring}} the Role of Financial Markets on Linkages} {Does foreign direct
  investment promote growth? {{Exploring}} the role of financial markets on
  linkages}.{\BBCQ}
\newblock

\PrintBackRefs{\CurrentBib}

\bibitem [\protect \citeauthoryear {%
Alsan%
\ \BBA {} Goldin%
}{%
Alsan%
\ \BBA {} Goldin%
}{%
{\protect \APACyear {2019}}%
}]{%
AlsanGoldin2019}
\APACinsertmetastar {%
AlsanGoldin2019}%
\begin{APACrefauthors}%
Alsan, M.%
\BCBT {}\ \BBA {} Goldin, C.%
\end{APACrefauthors}%
\unskip\
\newblock
\APACrefYearMonthDay{2019}{}{}.
\newblock
{\BBOQ}\APACrefatitle {Watersheds in Child Mortality: {{The}} Role of Effective
  Water and Sewerage Infrastructure, 1880\textendash 1920} {Watersheds in child
  mortality: {{The}} role of effective water and sewerage infrastructure,
  1880\textendash 1920}.{\BBCQ}
\newblock
\APACjournalVolNumPages{Journal of Political Economy}{127}{2}{586--638}.
\newblock
\begin{APACrefDOI} \doi{10/ghmc6z} \end{APACrefDOI}
\PrintBackRefs{\CurrentBib}

\bibitem [\protect \citeauthoryear {%
Amiti%
\ \BBA {} Konings%
}{%
Amiti%
\ \BBA {} Konings%
}{%
{\protect \APACyear {2007}}%
}]{%
AmitiKonings2007}
\APACinsertmetastar {%
AmitiKonings2007}%
\begin{APACrefauthors}%
Amiti, M.%
\BCBT {}\ \BBA {} Konings, J.%
\end{APACrefauthors}%
\unskip\
\newblock
\APACrefYearMonthDay{2007}{}{}.
\newblock
{\BBOQ}\APACrefatitle {Trade Liberalization, Intermediate Inputs, and
  Productivity: {{Evidence}} from Indonesia} {Trade liberalization,
  intermediate inputs, and productivity: {{Evidence}} from indonesia}.{\BBCQ}
\newblock
\APACjournalVolNumPages{American Economic Review}{97}{5}{1611--1638}.
\newblock
\begin{APACrefDOI} \doi{10/cqjs8g} \end{APACrefDOI}
\PrintBackRefs{\CurrentBib}

\bibitem [\protect \citeauthoryear {%
Arellano%
\ \BBA {} Bonhomme%
}{%
Arellano%
\ \BBA {} Bonhomme%
}{%
{\protect \APACyear {2012}}%
}]{%
arellanoIdentifyingDistributionalCharacteristics2012}
\APACinsertmetastar {%
arellanoIdentifyingDistributionalCharacteristics2012}%
\begin{APACrefauthors}%
Arellano, M.%
\BCBT {}\ \BBA {} Bonhomme, S.%
\end{APACrefauthors}%
\unskip\
\newblock
\APACrefYearMonthDay{2012}{}{}.
\newblock
{\BBOQ}\APACrefatitle {Identifying {{Distributional Characteristics}} in
  {{Random Coefficients Panel Data Models}}} {Identifying {{Distributional
  Characteristics}} in {{Random Coefficients Panel Data Models}}}.{\BBCQ}
\newblock

\PrintBackRefs{\CurrentBib}

\bibitem [\protect \citeauthoryear {%
Balli%
\ \BBA {} S{\o}rensen%
}{%
Balli%
\ \BBA {} S{\o}rensen%
}{%
{\protect \APACyear {2013}}%
}]{%
balliInteractionEffectsEconometrics2013}
\APACinsertmetastar {%
balliInteractionEffectsEconometrics2013}%
\begin{APACrefauthors}%
Balli, H\BPBI O.%
\BCBT {}\ \BBA {} S{\o}rensen, B\BPBI E.%
\end{APACrefauthors}%
\unskip\
\newblock
\APACrefYearMonthDay{2013}{}{}.
\newblock
{\BBOQ}\APACrefatitle {Interaction Effects in Econometrics} {Interaction
  effects in econometrics}.{\BBCQ}
\newblock
\APACjournalVolNumPages{Empirical Economics}{45}{1}{583--603}.
\newblock
\begin{APACrefDOI} \doi{10/gdjcbf} \end{APACrefDOI}
\PrintBackRefs{\CurrentBib}

\bibitem [\protect \citeauthoryear {%
Bekhtiar%
, Bittschi%
\BCBL {}\ \BBA {} Sellner%
}{%
Bekhtiar%
\ \protect \BOthers {.}}{%
{\protect \APACyear {2024}}%
}]{%
Bekhtiaretal2024}
\APACinsertmetastar {%
Bekhtiaretal2024}%
\begin{APACrefauthors}%
Bekhtiar, K.%
, Bittschi, B.%
\BCBL {}\ \BBA {} Sellner, R.%
\end{APACrefauthors}%
\unskip\
\newblock
\APACrefYearMonthDay{2024}{}{}.
\newblock
{\BBOQ}\APACrefatitle {Robots at work? Pitfalls of industry-level data} {Robots
  at work? pitfalls of industry-level data}.{\BBCQ}
\newblock
\APACjournalVolNumPages{Journal of Applied Econometrics}{39}{6}{1180-1189}.
\newblock
\begin{APACrefDOI} \doi{https://doi.org/10.1002/jae.3073} \end{APACrefDOI}
\PrintBackRefs{\CurrentBib}

\bibitem [\protect \citeauthoryear {%
Berman%
, Martin%
\BCBL {}\ \BBA {} Mayer%
}{%
Berman%
\ \protect \BOthers {.}}{%
{\protect \APACyear {2012}}%
}]{%
BermanMartinMayer2012}
\APACinsertmetastar {%
BermanMartinMayer2012}%
\begin{APACrefauthors}%
Berman, N.%
, Martin, P.%
\BCBL {}\ \BBA {} Mayer, T.%
\end{APACrefauthors}%
\unskip\
\newblock
\APACrefYearMonthDay{2012}{}{}.
\newblock
{\BBOQ}\APACrefatitle {How Do Different Exporters React to Exchange Rate
  Changes?} {How do different exporters react to exchange rate changes?}{\BBCQ}
\newblock
\APACjournalVolNumPages{Quarterly Journal of Economics}{127}{1}{437--492}.
\newblock
\begin{APACrefDOI} \doi{10/fx8562} \end{APACrefDOI}
\PrintBackRefs{\CurrentBib}

\bibitem [\protect \citeauthoryear {%
Bloom%
, Draca%
\BCBL {}\ \BBA {} Van~Reenen%
}{%
Bloom%
\ \protect \BOthers {.}}{%
{\protect \APACyear {2016}}%
}]{%
BloomDracaReenen2016}
\APACinsertmetastar {%
BloomDracaReenen2016}%
\begin{APACrefauthors}%
Bloom, N.%
, Draca, M.%
\BCBL {}\ \BBA {} Van~Reenen, J.%
\end{APACrefauthors}%
\unskip\
\newblock
\APACrefYearMonthDay{2016}{}{}.
\newblock
{\BBOQ}\APACrefatitle {Trade Induced Technical Change? {{The}} Impact of
  Chinese Imports on Innovation, {{IT}} and Productivity} {Trade induced
  technical change? {{The}} impact of chinese imports on innovation, {{IT}} and
  productivity}.{\BBCQ}
\newblock
\APACjournalVolNumPages{Review of Economic Studies}{83}{1}{87--117}.
\newblock
\begin{APACrefDOI} \doi{10/f8ntgw} \end{APACrefDOI}
\PrintBackRefs{\CurrentBib}

\bibitem [\protect \citeauthoryear {%
Bloom%
, Sadun%
\BCBL {}\ \BBA {} Van~Reenen%
}{%
Bloom%
\ \protect \BOthers {.}}{%
{\protect \APACyear {2012}}%
}]{%
BloomSadunReenen2012}
\APACinsertmetastar {%
BloomSadunReenen2012}%
\begin{APACrefauthors}%
Bloom, N.%
, Sadun, R.%
\BCBL {}\ \BBA {} Van~Reenen, J.%
\end{APACrefauthors}%
\unskip\
\newblock
\APACrefYearMonthDay{2012}{}{}.
\newblock
{\BBOQ}\APACrefatitle {Americans Do {{IT}} Better: {{US}} Multinationals and
  the Productivity Miracle} {Americans do {{IT}} better: {{US}} multinationals
  and the productivity miracle}.{\BBCQ}
\newblock
\APACjournalVolNumPages{American Economic Review}{102}{1}{167--201}.
\newblock
\begin{APACrefDOI} \doi{10/gftrv4} \end{APACrefDOI}
\PrintBackRefs{\CurrentBib}

\bibitem [\protect \citeauthoryear {%
Borensztein%
, Gregorio%
\BCBL {}\ \BBA {} Lee%
}{%
Borensztein%
\ \protect \BOthers {.}}{%
{\protect \APACyear {1998}}%
}]{%
borenszteinHowDoesForeign1998}
\APACinsertmetastar {%
borenszteinHowDoesForeign1998}%
\begin{APACrefauthors}%
Borensztein, E.%
, Gregorio, J\BPBI D.%
\BCBL {}\ \BBA {} Lee, J\BHBI W.%
\end{APACrefauthors}%
\unskip\
\newblock
\APACrefYearMonthDay{1998}{}{}.
\newblock
{\BBOQ}\APACrefatitle {How Does Foreign Direct Investment Affect Economic
  Growth?} {How does foreign direct investment affect economic growth?}{\BBCQ}
\newblock
\APACjournalVolNumPages{Journal of International Economics}{}{}{21}.
\newblock
\begin{APACrefDOI} \doi{10/d7n25r} \end{APACrefDOI}
\PrintBackRefs{\CurrentBib}

\bibitem [\protect \citeauthoryear {%
Burnside%
\ \BBA {} Dollar%
}{%
Burnside%
\ \BBA {} Dollar%
}{%
{\protect \APACyear {2000}}%
}]{%
burnsideAidPoliciesGrowth2000}
\APACinsertmetastar {%
burnsideAidPoliciesGrowth2000}%
\begin{APACrefauthors}%
Burnside, C.%
\BCBT {}\ \BBA {} Dollar, D.%
\end{APACrefauthors}%
\unskip\
\newblock
\APACrefYearMonthDay{2000}{}{}.
\newblock
{\BBOQ}\APACrefatitle {Aid, {{Policies}}, and {{Growth}}} {Aid, {{Policies}},
  and {{Growth}}}.{\BBCQ}
\newblock
\APACjournalVolNumPages{American Economic Review}{90}{4}{23}.
\newblock
\begin{APACrefDOI} \doi{10/dv7zrd} \end{APACrefDOI}
\PrintBackRefs{\CurrentBib}

\bibitem [\protect \citeauthoryear {%
Chamberlain%
}{%
Chamberlain%
}{%
{\protect \APACyear {1992}}%
}]{%
Chamberlain1992}
\APACinsertmetastar {%
Chamberlain1992}%
\begin{APACrefauthors}%
Chamberlain, G.%
\end{APACrefauthors}%
\unskip\
\newblock
\APACrefYearMonthDay{1992}{}{}.
\newblock
{\BBOQ}\APACrefatitle {Efficiency {{Bounds}} for {{Semiparametric Regression}}}
  {Efficiency {{Bounds}} for {{Semiparametric Regression}}}.{\BBCQ}
\newblock
\APACjournalVolNumPages{Econometrica}{60}{3}{567--596}.
\newblock
\begin{APACrefDOI} \doi{10/bq5bjf} \end{APACrefDOI}
\PrintBackRefs{\CurrentBib}

\bibitem [\protect \citeauthoryear {%
Chen%
\ \BBA {} Frey%
}{%
Chen%
\ \BBA {} Frey%
}{%
{\protect \APACyear {2024}}%
}]{%
ChenFrey2024}
\APACinsertmetastar {%
ChenFrey2024}%
\begin{APACrefauthors}%
Chen, C.%
\BCBT {}\ \BBA {} Frey, C\BPBI B.%
\end{APACrefauthors}%
\unskip\
\newblock
\APACrefYearMonthDay{2024}{}{}.
\newblock
{\BBOQ}\APACrefatitle {{Robots and reshoring: a comparative study of
  automation, trade, and employment in Europe}} {{Robots and reshoring: a
  comparative study of automation, trade, and employment in Europe}}.{\BBCQ}
\newblock
\APACjournalVolNumPages{Industrial and Corporate Change}{33}{6}{1331-1377}.
\PrintBackRefs{\CurrentBib}

\bibitem [\protect \citeauthoryear {%
Chetty%
, Hendren%
\BCBL {}\ \BBA {} Katz%
}{%
Chetty%
\ \protect \BOthers {.}}{%
{\protect \APACyear {2016}}%
}]{%
chettyEffectsExposureBetter2016a}
\APACinsertmetastar {%
chettyEffectsExposureBetter2016a}%
\begin{APACrefauthors}%
Chetty, R.%
, Hendren, N.%
\BCBL {}\ \BBA {} Katz, L\BPBI F.%
\end{APACrefauthors}%
\unskip\
\newblock
\APACrefYearMonthDay{2016}{{\APACmonth{04}}}{}.
\newblock
{\BBOQ}\APACrefatitle {The {{Effects}} of {{Exposure}} to {{Better
  Neighborhoods}} on {{Children}}: {{New Evidence}} from the {{Moving}} to
  {{Opportunity Experiment}}} {The {{Effects}} of {{Exposure}} to {{Better
  Neighborhoods}} on {{Children}}: {{New Evidence}} from the {{Moving}} to
  {{Opportunity Experiment}}}.{\BBCQ}
\newblock
\APACjournalVolNumPages{American Economic Review}{106}{4}{855--902}.
\newblock
\begin{APACrefDOI} \doi{10/f8gpmb} \end{APACrefDOI}
\PrintBackRefs{\CurrentBib}

\bibitem [\protect \citeauthoryear {%
Couttenier%
, Petrencu%
, Rohner%
\BCBL {}\ \BBA {} Thoenig%
}{%
Couttenier%
\ \protect \BOthers {.}}{%
{\protect \APACyear {2019}}%
}]{%
couttenierViolentLegacyConflict2019a}
\APACinsertmetastar {%
couttenierViolentLegacyConflict2019a}%
\begin{APACrefauthors}%
Couttenier, M.%
, Petrencu, V.%
, Rohner, D.%
\BCBL {}\ \BBA {} Thoenig, M.%
\end{APACrefauthors}%
\unskip\
\newblock
\APACrefYearMonthDay{2019}{{\APACmonth{12}}}{}.
\newblock
{\BBOQ}\APACrefatitle {The {{Violent Legacy}} of {{Conflict}}: {{Evidence}} on
  {{Asylum Seekers}}, {{Crime}}, and {{Public Policy}} in {{Switzerland}}} {The
  {{Violent Legacy}} of {{Conflict}}: {{Evidence}} on {{Asylum Seekers}},
  {{Crime}}, and {{Public Policy}} in {{Switzerland}}}.{\BBCQ}
\newblock
\APACjournalVolNumPages{American Economic Review}{109}{12}{4378--4425}.
\newblock
\begin{APACrefDOI} \doi{10.1257/aer.20170263} \end{APACrefDOI}
\PrintBackRefs{\CurrentBib}

\bibitem [\protect \citeauthoryear {%
Dauth%
, Findeisen%
, Suedekum%
\BCBL {}\ \BBA {} Woessner%
}{%
Dauth%
\ \protect \BOthers {.}}{%
{\protect \APACyear {2021}}%
}]{%
Dauthetal2021}
\APACinsertmetastar {%
Dauthetal2021}%
\begin{APACrefauthors}%
Dauth, W.%
, Findeisen, S.%
, Suedekum, J.%
\BCBL {}\ \BBA {} Woessner, N.%
\end{APACrefauthors}%
\unskip\
\newblock
\APACrefYearMonthDay{2021}{}{}.
\newblock
{\BBOQ}\APACrefatitle {{The Adjustment of Labor Markets to Robots}} {{The
  Adjustment of Labor Markets to Robots}}.{\BBCQ}
\newblock
\APACjournalVolNumPages{Journal of the European Economic
  Association}{19}{6}{3104-3153}.
\PrintBackRefs{\CurrentBib}

\bibitem [\protect \citeauthoryear {%
{de Chaisemartin}%
\ \BBA {} D'Haultf{\oe}uille%
}{%
{de Chaisemartin}%
\ \BBA {} D'Haultf{\oe}uille%
}{%
{\protect \APACyear {2022}}%
}]{%
dechaisemartinTwoWayFixedEffects2022a}
\APACinsertmetastar {%
dechaisemartinTwoWayFixedEffects2022a}%
\begin{APACrefauthors}%
{de Chaisemartin}, C.%
\BCBT {}\ \BBA {} D'Haultf{\oe}uille, X.%
\end{APACrefauthors}%
\unskip\
\newblock
\APACrefYearMonthDay{2022}{}{}.
\newblock
\APACrefbtitle {Two-{{Way Fixed Effects}} and {{Differences-in-Differences}}
  with {{Heterogeneous Treatment Effects}}: {{A Survey}}} {Two-{{Way Fixed
  Effects}} and {{Differences-in-Differences}} with {{Heterogeneous Treatment
  Effects}}: {{A Survey}}}\ \APACbVolEdTR {}{Working Paper\ \BNUM\ 29691}.
\newblock
\APACaddressInstitution{}{National Bureau of Economic Research}.
\PrintBackRefs{\CurrentBib}

\bibitem [\protect \citeauthoryear {%
{de Chaisemartin}%
\ \BBA {} D'Haultf{\oe}uille%
}{%
{de Chaisemartin}%
\ \BBA {} D'Haultf{\oe}uille%
}{%
{\protect \APACyear {2024}}%
}]{%
dechaisemartinCredibleAnswersHard2024}
\APACinsertmetastar {%
dechaisemartinCredibleAnswersHard2024}%
\begin{APACrefauthors}%
{de Chaisemartin}, C.%
\BCBT {}\ \BBA {} D'Haultf{\oe}uille, X.%
\end{APACrefauthors}%
\unskip\
\newblock
\APACrefYearMonthDay{2024}{}{}.
\newblock
{\BBOQ}\APACrefatitle {Credible {{Answers}} to {{Hard Questions}}:
  {{Differences-in-Differences}} for {{Natural Experiments}}} {Credible
  {{Answers}} to {{Hard Questions}}: {{Differences-in-Differences}} for
  {{Natural Experiments}}}.{\BBCQ}
\newblock

\PrintBackRefs{\CurrentBib}

\bibitem [\protect \citeauthoryear {%
{de Vries}%
, Gentile%
, Miroudot%
\BCBL {}\ \BBA {} Wacker%
}{%
{de Vries}%
\ \protect \BOthers {.}}{%
{\protect \APACyear {2020}}%
}]{%
deVriesetal2020}
\APACinsertmetastar {%
deVriesetal2020}%
\begin{APACrefauthors}%
{de Vries}, G\BPBI J.%
, Gentile, E.%
, Miroudot, S.%
\BCBL {}\ \BBA {} Wacker, K\BPBI M.%
\end{APACrefauthors}%
\unskip\
\newblock
\APACrefYearMonthDay{2020}{}{}.
\newblock
{\BBOQ}\APACrefatitle {The rise of robots and the fall of routine jobs} {The
  rise of robots and the fall of routine jobs}.{\BBCQ}
\newblock
\APACjournalVolNumPages{Labour Economics}{66}{}{101885}.
\newblock
\begin{APACrefDOI} \doi{https://doi.org/10.1016/j.labeco.2020.101885}
  \end{APACrefDOI}
\PrintBackRefs{\CurrentBib}

\bibitem [\protect \citeauthoryear {%
de Chaisemartin%
\ \BBA {} Lei%
}{%
de Chaisemartin%
\ \BBA {} Lei%
}{%
{\protect \APACyear {2023}}%
}]{%
chaisemartinMoreRobustEstimators2023}
\APACinsertmetastar {%
chaisemartinMoreRobustEstimators2023}%
\begin{APACrefauthors}%
de Chaisemartin, C.%
\BCBT {}\ \BBA {} Lei, Z.%
\end{APACrefauthors}%
\unskip\
\newblock
\APACrefYearMonthDay{2023}{{\APACmonth{09}}}{}.
\newblock
\APACrefbtitle {More {{Robust Estimators}} for {{Instrumental-Variable Panel
  Designs}}, {{With An Application}} to the {{Effect}} of {{Imports}} from
  {{China}} on {{US Employment}}} {More {{Robust Estimators}} for
  {{Instrumental-Variable Panel Designs}}, {{With An Application}} to the
  {{Effect}} of {{Imports}} from {{China}} on {{US Employment}}}\ (\BNUM\
  arXiv:2103.06437).
\newblock
\APACaddressPublisher{}{arXiv}.
\newblock
\begin{APACrefDOI} \doi{10.48550/arXiv.2103.06437} \end{APACrefDOI}
\PrintBackRefs{\CurrentBib}

\bibitem [\protect \citeauthoryear {%
Dijkstra%
\ \BBA {} Wacker%
}{%
Dijkstra%
\ \BBA {} Wacker%
}{%
{\protect \APACyear {2025}}%
}]{%
DijkstraWacker2025}
\APACinsertmetastar {%
DijkstraWacker2025}%
\begin{APACrefauthors}%
Dijkstra, H.%
\BCBT {}\ \BBA {} Wacker, K\BPBI M.%
\end{APACrefauthors}%
\unskip\
\newblock
\APACrefYearMonthDay{2025}{}{}.
\newblock
\APACrefbtitle {Robots, Shoring Patterns, and Employment} {Robots, shoring
  patterns, and employment}\ \APACbVolEdTR {}{Working Paper\ \BNUM\
  forthcoming}.
\newblock
\begin{APACrefURL}
  \url{https://drive.google.com/file/d/1wA8JcEmc9TROAe7Pr0_kv1WvD57OusRv/view}
  \end{APACrefURL}
\PrintBackRefs{\CurrentBib}

\bibitem [\protect \citeauthoryear {%
Duflo%
, Dupas%
\BCBL {}\ \BBA {} Kremer%
}{%
Duflo%
\ \protect \BOthers {.}}{%
{\protect \APACyear {2011}}%
}]{%
dufloPeerEffectsTeacher2011}
\APACinsertmetastar {%
dufloPeerEffectsTeacher2011}%
\begin{APACrefauthors}%
Duflo, E.%
, Dupas, P.%
\BCBL {}\ \BBA {} Kremer, M.%
\end{APACrefauthors}%
\unskip\
\newblock
\APACrefYearMonthDay{2011}{}{}.
\newblock
{\BBOQ}\APACrefatitle {Peer {{Effects}}, {{Teacher Incentives}}, and the
  {{Impact}} of {{Tracking}}: {{Evidence}} from a {{Randomized Evaluation}} in
  {{Kenya}}} {Peer {{Effects}}, {{Teacher Incentives}}, and the {{Impact}} of
  {{Tracking}}: {{Evidence}} from a {{Randomized Evaluation}} in
  {{Kenya}}}.{\BBCQ}
\newblock

\PrintBackRefs{\CurrentBib}

\bibitem [\protect \citeauthoryear {%
Durham%
}{%
Durham%
}{%
{\protect \APACyear {2004}}%
}]{%
durham04}
\APACinsertmetastar {%
durham04}%
\begin{APACrefauthors}%
Durham, J\BPBI B.%
\end{APACrefauthors}%
\unskip\
\newblock
\APACrefYearMonthDay{2004}{}{}.
\newblock
{\BBOQ}\APACrefatitle {Absorptive capacity and the effects of foreign direct
  investment and equity foreign portfolio investment on economic growth}
  {Absorptive capacity and the effects of foreign direct investment and equity
  foreign portfolio investment on economic growth}.{\BBCQ}
\newblock
\APACjournalVolNumPages{European Economic Review}{48}{2}{285-306}.
\PrintBackRefs{\CurrentBib}

\bibitem [\protect \citeauthoryear {%
Epifani%
\ \BBA {} Gancia%
}{%
Epifani%
\ \BBA {} Gancia%
}{%
{\protect \APACyear {2009}}%
}]{%
EpifaniGancia2009}
\APACinsertmetastar {%
EpifaniGancia2009}%
\begin{APACrefauthors}%
Epifani, P.%
\BCBT {}\ \BBA {} Gancia, G.%
\end{APACrefauthors}%
\unskip\
\newblock
\APACrefYearMonthDay{2009}{}{}.
\newblock
{\BBOQ}\APACrefatitle {Openness, Government Size and the Terms of Trade}
  {Openness, government size and the terms of trade}.{\BBCQ}
\newblock
\APACjournalVolNumPages{The Review of Economic Studies}{76}{2}{629--668}.
\newblock
\begin{APACrefDOI} \doi{10/dqhf3d} \end{APACrefDOI}
\PrintBackRefs{\CurrentBib}

\bibitem [\protect \citeauthoryear {%
Feenstra%
, Inklaar%
\BCBL {}\ \BBA {} Timmer%
}{%
Feenstra%
\ \protect \BOthers {.}}{%
{\protect \APACyear {2015}}%
}]{%
feenstrapwt9}
\APACinsertmetastar {%
feenstrapwt9}%
\begin{APACrefauthors}%
Feenstra, R\BPBI C.%
, Inklaar, R.%
\BCBL {}\ \BBA {} Timmer, M\BPBI P.%
\end{APACrefauthors}%
\unskip\
\newblock
\APACrefYearMonthDay{2015}{}{}.
\newblock
{\BBOQ}\APACrefatitle {The Next Generation of the Penn World Table} {The next
  generation of the penn world table}.{\BBCQ}
\newblock
\APACjournalVolNumPages{American Economic Review}{105}{10}{3150-82}.
\PrintBackRefs{\CurrentBib}

\bibitem [\protect \citeauthoryear {%
Fern{\'a}ndez-Val%
, Gao%
, Liao%
\BCBL {}\ \BBA {} Vella%
}{%
Fern{\'a}ndez-Val%
\ \protect \BOthers {.}}{%
{\protect \APACyear {2022}}%
}]{%
fernandez2022dynamic}
\APACinsertmetastar {%
fernandez2022dynamic}%
\begin{APACrefauthors}%
Fern{\'a}ndez-Val, I.%
, Gao, W\BPBI Y.%
, Liao, Y.%
\BCBL {}\ \BBA {} Vella, F.%
\end{APACrefauthors}%
\unskip\
\newblock
\APACrefYearMonthDay{2022}{}{}.
\newblock
{\BBOQ}\APACrefatitle {Dynamic heterogeneous distribution regression panel
  models, with an application to labor income processes} {Dynamic heterogeneous
  distribution regression panel models, with an application to labor income
  processes}.{\BBCQ}
\newblock
\APACjournalVolNumPages{arXiv preprint arXiv:2202.04154}{}{}{}.
\PrintBackRefs{\CurrentBib}

\bibitem [\protect \citeauthoryear {%
Fern\'{a}ndez-Val%
\ \BBA {} Lee%
}{%
Fern\'{a}ndez-Val%
\ \BBA {} Lee%
}{%
{\protect \APACyear {2013}}%
}]{%
fernandezvalNonadditiveUnobservedHeterogeneity2013a}
\APACinsertmetastar {%
fernandezvalNonadditiveUnobservedHeterogeneity2013a}%
\begin{APACrefauthors}%
Fern\'{a}ndez-Val, I.%
\BCBT {}\ \BBA {} Lee, J.%
\end{APACrefauthors}%
\unskip\
\newblock
\APACrefYearMonthDay{2013}{}{}.
\newblock
{\BBOQ}\APACrefatitle {Panel data models with nonadditive unobserved
  heterogeneity: Estimation and inference} {Panel data models with nonadditive
  unobserved heterogeneity: Estimation and inference}.{\BBCQ}
\newblock
\APACjournalVolNumPages{Quantitative Economics}{4}{3}{453-481}.
\newblock
\begin{APACrefDOI} \doi{https://doi.org/10.3982/QE75} \end{APACrefDOI}
\PrintBackRefs{\CurrentBib}

\bibitem [\protect \citeauthoryear {%
Giesselmann%
\ \BBA {} {Schmidt-Catran}%
}{%
Giesselmann%
\ \BBA {} {Schmidt-Catran}%
}{%
{\protect \APACyear {2020}}%
}]{%
giesselmannInteractionsFixedEffects2020}
\APACinsertmetastar {%
giesselmannInteractionsFixedEffects2020}%
\begin{APACrefauthors}%
Giesselmann, M.%
\BCBT {}\ \BBA {} {Schmidt-Catran}, A\BPBI W.%
\end{APACrefauthors}%
\unskip\
\newblock
\APACrefYearMonthDay{2020}{}{}.
\newblock
{\BBOQ}\APACrefatitle {Interactions in {{Fixed Effects Regression Models}}}
  {Interactions in {{Fixed Effects Regression Models}}}.{\BBCQ}
\newblock

\PrintBackRefs{\CurrentBib}

\bibitem [\protect \citeauthoryear {%
Graetz%
\ \BBA {} Michaels%
}{%
Graetz%
\ \BBA {} Michaels%
}{%
{\protect \APACyear {2018}}%
}]{%
GraetzMichaels2018}
\APACinsertmetastar {%
GraetzMichaels2018}%
\begin{APACrefauthors}%
Graetz, G.%
\BCBT {}\ \BBA {} Michaels, G.%
\end{APACrefauthors}%
\unskip\
\newblock
\APACrefYearMonthDay{2018}{12}{}.
\newblock
{\BBOQ}\APACrefatitle {{Robots at Work}} {{Robots at Work}}.{\BBCQ}
\newblock
\APACjournalVolNumPages{The Review of Economics and
  Statistics}{100}{5}{753-768}.
\PrintBackRefs{\CurrentBib}

\bibitem [\protect \citeauthoryear {%
Graham%
\ \BBA {} Powell%
}{%
Graham%
\ \BBA {} Powell%
}{%
{\protect \APACyear {2012}}%
}]{%
grahamIdentificationEstimationAverage2012}
\APACinsertmetastar {%
grahamIdentificationEstimationAverage2012}%
\begin{APACrefauthors}%
Graham, B\BPBI S.%
\BCBT {}\ \BBA {} Powell, J\BPBI L.%
\end{APACrefauthors}%
\unskip\
\newblock
\APACrefYearMonthDay{2012}{}{}.
\newblock
{\BBOQ}\APACrefatitle {Identification and {{Estimation}} of {{Average Partial
  Effects}} in "{{Irregular}}" {{Correlated Random Coefficient Panel Data
  Models}}} {Identification and {{Estimation}} of {{Average Partial Effects}}
  in "{{Irregular}}" {{Correlated Random Coefficient Panel Data
  Models}}}.{\BBCQ}
\newblock
\APACjournalVolNumPages{Econometrica}{80}{5}{2105--2152}.
\newblock
\begin{APACrefDOI} \doi{10.3982/ECTA8220} \end{APACrefDOI}
\PrintBackRefs{\CurrentBib}

\bibitem [\protect \citeauthoryear {%
Guarascio%
, Piccirillo%
\BCBL {}\ \BBA {} Reljic%
}{%
Guarascio%
\ \protect \BOthers {.}}{%
{\protect \APACyear {2024}}%
}]{%
Guarascioetal2024}
\APACinsertmetastar {%
Guarascioetal2024}%
\begin{APACrefauthors}%
Guarascio, D.%
, Piccirillo, A.%
\BCBL {}\ \BBA {} Reljic, J.%
\end{APACrefauthors}%
\unskip\
\newblock
\APACrefYearMonthDay{2024}{}{}.
\newblock
\APACrefbtitle {{Will robot replace workers? Assessing the impact of robots on
  employment and wages with meta-analysis}} {{Will robot replace workers?
  Assessing the impact of robots on employment and wages with meta-analysis}}\
  \APACbVolEdTR{}{\BTR{}}.
\PrintBackRefs{\CurrentBib}

\bibitem [\protect \citeauthoryear {%
Herrera%
, Ordo\~nez%
\BCBL {}\ \BBA {} Trebesch%
}{%
Herrera%
\ \protect \BOthers {.}}{%
{\protect \APACyear {2020}}%
}]{%
HerreraTrebesch2020}
\APACinsertmetastar {%
HerreraTrebesch2020}%
\begin{APACrefauthors}%
Herrera, H.%
, Ordo\~nez, G.%
\BCBL {}\ \BBA {} Trebesch, C.%
\end{APACrefauthors}%
\unskip\
\newblock
\APACrefYearMonthDay{2020}{}{}.
\newblock
{\BBOQ}\APACrefatitle {Political Booms, Financial Crises} {Political booms,
  financial crises}.{\BBCQ}
\newblock
\APACjournalVolNumPages{Journal of Political Economy}{128}{2}{507--543}.
\newblock
\begin{APACrefDOI} \doi{10/ggfxc3} \end{APACrefDOI}
\PrintBackRefs{\CurrentBib}

\bibitem [\protect \citeauthoryear {%
Jurat%
, Klump%
\BCBL {}\ \BBA {} Schneider%
}{%
Jurat%
\ \protect \BOthers {.}}{%
{\protect \APACyear {2023}}%
}]{%
Juratetal2023}
\APACinsertmetastar {%
Juratetal2023}%
\begin{APACrefauthors}%
Jurat, A.%
, Klump, R.%
\BCBL {}\ \BBA {} Schneider, F.%
\end{APACrefauthors}%
\unskip\
\newblock
\APACrefYearMonthDay{2023}{}{}.
\newblock
\APACrefbtitle {Robots and Wages: A Meta-Analysis} {Robots and wages: A
  meta-analysis}\ \APACbVolEdTR{}{\BTR{}}.
\newblock
\APACaddressInstitution{}{ZBW - Leibniz Information Centre for Economics, Kiel,
  Hamburg}.
\PrintBackRefs{\CurrentBib}

\bibitem [\protect \citeauthoryear {%
Laage%
}{%
Laage%
}{%
{\protect \APACyear {2020}}%
}]{%
laageCorrelatedRandomCoefficient2020b}
\APACinsertmetastar {%
laageCorrelatedRandomCoefficient2020b}%
\begin{APACrefauthors}%
Laage, L.%
\end{APACrefauthors}%
\unskip\
\newblock
\APACrefYearMonthDay{2020}{{\APACmonth{03}}}{}.
\newblock
\APACrefbtitle {A {{Correlated Random Coefficient Panel Model}} with
  {{Time-Varying Endogeneity}}} {A {{Correlated Random Coefficient Panel
  Model}} with {{Time-Varying Endogeneity}}}\ (\BNUM\ arXiv:2003.09367).
\newblock
\APAChowpublished {arXiv:2003.09367}.
\newblock
\APACaddressPublisher{}{{arXiv}}.
\newblock
\begin{APACrefDOI} \doi{10.48550/arXiv.2003.09367} \end{APACrefDOI}
\PrintBackRefs{\CurrentBib}

\bibitem [\protect \citeauthoryear {%
Laage%
}{%
Laage%
}{%
{\protect \APACyear {2024}}%
}]{%
laage2024crcpublished}
\APACinsertmetastar {%
laage2024crcpublished}%
\begin{APACrefauthors}%
Laage, L.%
\end{APACrefauthors}%
\unskip\
\newblock
\APACrefYearMonthDay{2024}{}{}.
\newblock
{\BBOQ}\APACrefatitle {A {{Correlated Random Coefficient Panel Model}} with
  {{Time-Varying Endogeneity}}} {A {{Correlated Random Coefficient Panel
  Model}} with {{Time-Varying Endogeneity}}}.{\BBCQ}
\newblock
\APACjournalVolNumPages{Journal of Econometrics}{242}{2}{105804}.
\PrintBackRefs{\CurrentBib}

\bibitem [\protect \citeauthoryear {%
List%
\ \BBA {} Sturm%
}{%
List%
\ \BBA {} Sturm%
}{%
{\protect \APACyear {2006}}%
}]{%
ListStrum2006}
\APACinsertmetastar {%
ListStrum2006}%
\begin{APACrefauthors}%
List, J\BPBI A.%
\BCBT {}\ \BBA {} Sturm, D\BPBI M.%
\end{APACrefauthors}%
\unskip\
\newblock
\APACrefYearMonthDay{2006}{}{}.
\newblock
{\BBOQ}\APACrefatitle {How Elections Matter: {{Theory}} and Evidence from
  Environmental Policy} {How elections matter: {{Theory}} and evidence from
  environmental policy}.{\BBCQ}
\newblock
\APACjournalVolNumPages{Quarterly Journal of Economics}{121}{4}{1249--1281}.
\PrintBackRefs{\CurrentBib}

\bibitem [\protect \citeauthoryear {%
MaCurdy%
}{%
MaCurdy%
}{%
{\protect \APACyear {1981}}%
}]{%
macurdyEmpiricalModelLabor1981}
\APACinsertmetastar {%
macurdyEmpiricalModelLabor1981}%
\begin{APACrefauthors}%
MaCurdy, T\BPBI E.%
\end{APACrefauthors}%
\unskip\
\newblock
\APACrefYearMonthDay{1981}{}{}.
\newblock
{\BBOQ}\APACrefatitle {An {{Empirical Model}} of {{Labor Supply}} in a
  {{Life-Cycle Setting}}} {An {{Empirical Model}} of {{Labor Supply}} in a
  {{Life-Cycle Setting}}}.{\BBCQ}
\newblock

\PrintBackRefs{\CurrentBib}

\bibitem [\protect \citeauthoryear {%
Manacorda%
\ \BBA {} Tesei%
}{%
Manacorda%
\ \BBA {} Tesei%
}{%
{\protect \APACyear {2020}}%
}]{%
manacordaLiberationTechnologyMobile2020}
\APACinsertmetastar {%
manacordaLiberationTechnologyMobile2020}%
\begin{APACrefauthors}%
Manacorda, M.%
\BCBT {}\ \BBA {} Tesei, A.%
\end{APACrefauthors}%
\unskip\
\newblock
\APACrefYearMonthDay{2020}{}{}.
\newblock
{\BBOQ}\APACrefatitle {Liberation {{Technology}}: {{Mobile Phones}} and
  {{Political Mobilization}} in {{Africa}}} {Liberation {{Technology}}:
  {{Mobile Phones}} and {{Political Mobilization}} in {{Africa}}}.{\BBCQ}
\newblock
\APACjournalVolNumPages{Econometrica}{88}{2}{533--567}.
\newblock
\begin{APACrefDOI} \doi{10.3982/ECTA14392} \end{APACrefDOI}
\PrintBackRefs{\CurrentBib}

\bibitem [\protect \citeauthoryear {%
Mondolo%
}{%
Mondolo%
}{%
{\protect \APACyear {2022}}%
}]{%
Mondolo2022}
\APACinsertmetastar {%
Mondolo2022}%
\begin{APACrefauthors}%
Mondolo, J.%
\end{APACrefauthors}%
\unskip\
\newblock
\APACrefYearMonthDay{2022}{}{}.
\newblock
{\BBOQ}\APACrefatitle {The composite link between technological change and
  employment: A survey of the literature} {The composite link between
  technological change and employment: A survey of the literature}.{\BBCQ}
\newblock
\APACjournalVolNumPages{Journal of Economic Surveys}{36}{4}{1027-1068}.
\newblock
\begin{APACrefDOI} \doi{https://doi.org/10.1111/joes.12469} \end{APACrefDOI}
\PrintBackRefs{\CurrentBib}

\bibitem [\protect \citeauthoryear {%
Muris%
\ \BBA {} Wacker%
}{%
Muris%
\ \BBA {} Wacker%
}{%
{\protect \APACyear {2022}}%
}]{%
MurisWacker2022}
\APACinsertmetastar {%
MurisWacker2022}%
\begin{APACrefauthors}%
Muris, C.%
\BCBT {}\ \BBA {} Wacker, K\BPBI M.%
\end{APACrefauthors}%
\unskip\
\newblock
\APACrefYearMonthDay{2022}{}{}.
\newblock
\APACrefbtitle {Estimating interaction effects with panel data} {Estimating
  interaction effects with panel data}\ \APACbVolEdTR{}{\BTR{}\ \BNUM\
  arXiv:2211.01557 [econ.EM]}.
\PrintBackRefs{\CurrentBib}

\bibitem [\protect \citeauthoryear {%
Murtazashvili%
\ \BBA {} Wooldridge%
}{%
Murtazashvili%
\ \BBA {} Wooldridge%
}{%
{\protect \APACyear {2008}}%
}]{%
murtazashviliFixedEffectsInstrumental2008}
\APACinsertmetastar {%
murtazashviliFixedEffectsInstrumental2008}%
\begin{APACrefauthors}%
Murtazashvili, I.%
\BCBT {}\ \BBA {} Wooldridge, J\BPBI M.%
\end{APACrefauthors}%
\unskip\
\newblock
\APACrefYearMonthDay{2008}{{\APACmonth{01}}}{}.
\newblock
{\BBOQ}\APACrefatitle {Fixed Effects Instrumental Variables Estimation in
  Correlated Random Coefficient Panel Data Models} {Fixed effects instrumental
  variables estimation in correlated random coefficient panel data
  models}.{\BBCQ}
\newblock
\APACjournalVolNumPages{Journal of Econometrics}{142}{1}{539--552}.
\newblock
\begin{APACrefDOI} \doi{10/b46wcn} \end{APACrefDOI}
\PrintBackRefs{\CurrentBib}

\bibitem [\protect \citeauthoryear {%
Norris%
, Pecenco%
\BCBL {}\ \BBA {} Weaver%
}{%
Norris%
\ \protect \BOthers {.}}{%
{\protect \APACyear {2021}}%
}]{%
norrisEffectsParentalSibling2021}
\APACinsertmetastar {%
norrisEffectsParentalSibling2021}%
\begin{APACrefauthors}%
Norris, S.%
, Pecenco, M.%
\BCBL {}\ \BBA {} Weaver, J.%
\end{APACrefauthors}%
\unskip\
\newblock
\APACrefYearMonthDay{2021}{}{}.
\newblock
{\BBOQ}\APACrefatitle {The {{Effects}} of {{Parental}} and {{Sibling
  Incarceration}}: {{Evidence}} from {{Ohio}}} {The {{Effects}} of {{Parental}}
  and {{Sibling Incarceration}}: {{Evidence}} from {{Ohio}}}.{\BBCQ}
\newblock

\PrintBackRefs{\CurrentBib}

\bibitem [\protect \citeauthoryear {%
Patel%
, Sandefur%
\BCBL {}\ \BBA {} Subramanian%
}{%
Patel%
\ \protect \BOthers {.}}{%
{\protect \APACyear {2021}}%
}]{%
pateletal2021}
\APACinsertmetastar {%
pateletal2021}%
\begin{APACrefauthors}%
Patel, D.%
, Sandefur, J.%
\BCBL {}\ \BBA {} Subramanian, A.%
\end{APACrefauthors}%
\unskip\
\newblock
\APACrefYearMonthDay{2021}{}{}.
\newblock
{\BBOQ}\APACrefatitle {The new era of unconditional convergence} {The new era
  of unconditional convergence}.{\BBCQ}
\newblock
\APACjournalVolNumPages{Journal of Development Economics}{152}{}{102687}.
\PrintBackRefs{\CurrentBib}

\bibitem [\protect \citeauthoryear {%
Pritchett%
\ \BBA {} Summers%
}{%
Pritchett%
\ \BBA {} Summers%
}{%
{\protect \APACyear {2014}}%
}]{%
PritchettSummers2014}
\APACinsertmetastar {%
PritchettSummers2014}%
\begin{APACrefauthors}%
Pritchett, L.%
\BCBT {}\ \BBA {} Summers, L\BPBI H.%
\end{APACrefauthors}%
\unskip\
\newblock
\APACrefYearMonthDay{2014}{}{}.
\newblock
\APACrefbtitle {Asiaphoria Meets Regression to the Mean} {Asiaphoria meets
  regression to the mean}\ \APACbVolEdTR {}{NBER Working Paper\ \BNUM\ 20573}.
\newblock
\APACaddressInstitution{}{National Bureau of Economic Research}.
\PrintBackRefs{\CurrentBib}

\bibitem [\protect \citeauthoryear {%
Reljic%
, Cirillo%
\BCBL {}\ \BBA {} Guarascio%
}{%
Reljic%
\ \protect \BOthers {.}}{%
{\protect \APACyear {2023}}%
}]{%
Reljicetal2023}
\APACinsertmetastar {%
Reljicetal2023}%
\begin{APACrefauthors}%
Reljic, J.%
, Cirillo, V.%
\BCBL {}\ \BBA {} Guarascio, D.%
\end{APACrefauthors}%
\unskip\
\newblock
\APACrefYearMonthDay{2023}{}{}.
\newblock
{\BBOQ}\APACrefatitle {{Regimes of robotization in Europe}} {{Regimes of
  robotization in Europe}}.{\BBCQ}
\newblock
\APACjournalVolNumPages{Economics Letters}{232}{}{111320}.
\PrintBackRefs{\CurrentBib}

\bibitem [\protect \citeauthoryear {%
Restrepo%
}{%
Restrepo%
}{%
{\protect \APACyear {2024}}%
}]{%
Restrepo2023}
\APACinsertmetastar {%
Restrepo2023}%
\begin{APACrefauthors}%
Restrepo, P.%
\end{APACrefauthors}%
\unskip\
\newblock
\APACrefYearMonthDay{2024}{}{}.
\newblock
{\BBOQ}\APACrefatitle {Automation: Theory, Evidence, and Outlook} {Automation:
  Theory, evidence, and outlook}.{\BBCQ}
\newblock
\APACjournalVolNumPages{Annual Review of Economics}{16}{}{1-25}.
\newblock
\begin{APACrefDOI}
  \doi{https://doi.org/10.1146/annurev-economics-090523-113355}
  \end{APACrefDOI}
\PrintBackRefs{\CurrentBib}

\bibitem [\protect \citeauthoryear {%
Sasaki%
\ \BBA {} Ura%
}{%
Sasaki%
\ \BBA {} Ura%
}{%
{\protect \APACyear {2021}}%
}]{%
sasakiSlowMoversPanel2021a}
\APACinsertmetastar {%
sasakiSlowMoversPanel2021a}%
\begin{APACrefauthors}%
Sasaki, Y.%
\BCBT {}\ \BBA {} Ura, T.%
\end{APACrefauthors}%
\unskip\
\newblock
\APACrefYearMonthDay{2021}{{\APACmonth{10}}}{}.
\newblock
\APACrefbtitle {Slow {{Movers}} in {{Panel Data}}} {Slow {{Movers}} in {{Panel
  Data}}}\ (\BNUM\ arXiv:2110.12041).
\newblock
\APAChowpublished {arXiv:2110.12041}.
\newblock
\APACaddressPublisher{}{{arXiv}}.
\PrintBackRefs{\CurrentBib}

\bibitem [\protect \citeauthoryear {%
Shambaugh%
}{%
Shambaugh%
}{%
{\protect \APACyear {2004}}%
}]{%
Shambaugh2004}
\APACinsertmetastar {%
Shambaugh2004}%
\begin{APACrefauthors}%
Shambaugh, J\BPBI C.%
\end{APACrefauthors}%
\unskip\
\newblock
\APACrefYearMonthDay{2004}{}{}.
\newblock
{\BBOQ}\APACrefatitle {The Effect of Fixed Exchange Rates on Monetary Policy}
  {The effect of fixed exchange rates on monetary policy}.{\BBCQ}
\newblock
\APACjournalVolNumPages{Quarterly Journal of Economics}{119}{1}{301--352}.
\newblock
\begin{APACrefDOI} \doi{10/dzhq6n} \end{APACrefDOI}
\PrintBackRefs{\CurrentBib}

\bibitem [\protect \citeauthoryear {%
S{\l}oczy{\'n}ski%
}{%
S{\l}oczy{\'n}ski%
}{%
{\protect \APACyear {2022}}%
}]{%
sloczynskiInterpretingOLSEstimands2022}
\APACinsertmetastar {%
sloczynskiInterpretingOLSEstimands2022}%
\begin{APACrefauthors}%
S{\l}oczy{\'n}ski, T.%
\end{APACrefauthors}%
\unskip\
\newblock
\APACrefYearMonthDay{2022}{{\APACmonth{05}}}{}.
\newblock
{\BBOQ}\APACrefatitle {Interpreting {{OLS Estimands When Treatment Effects Are
  Heterogeneous}}: {{Smaller Groups Get Larger Weights}}} {Interpreting {{OLS
  Estimands When Treatment Effects Are Heterogeneous}}: {{Smaller Groups Get
  Larger Weights}}}.{\BBCQ}
\newblock
\APACjournalVolNumPages{The Review of Economics and
  Statistics}{104}{3}{501--509}.
\newblock
\begin{APACrefDOI} \doi{10.1162/rest_a_00953} \end{APACrefDOI}
\PrintBackRefs{\CurrentBib}

\bibitem [\protect \citeauthoryear {%
Spilimbergo%
}{%
Spilimbergo%
}{%
{\protect \APACyear {2009}}%
}]{%
spilimbergoDemocracyForeignEducation2009}
\APACinsertmetastar {%
spilimbergoDemocracyForeignEducation2009}%
\begin{APACrefauthors}%
Spilimbergo, A.%
\end{APACrefauthors}%
\unskip\
\newblock
\APACrefYearMonthDay{2009}{}{}.
\newblock
{\BBOQ}\APACrefatitle {Democracy and {{Foreign Education}}} {Democracy and
  {{Foreign Education}}}.{\BBCQ}
\newblock

\PrintBackRefs{\CurrentBib}

\bibitem [\protect \citeauthoryear {%
Stock%
\ \BBA {} Watson%
}{%
Stock%
\ \BBA {} Watson%
}{%
{\protect \APACyear {2015}}%
}]{%
stockIntroductionEconometrics2015}
\APACinsertmetastar {%
stockIntroductionEconometrics2015}%
\begin{APACrefauthors}%
Stock, J\BPBI H.%
\BCBT {}\ \BBA {} Watson, M\BPBI W.%
\end{APACrefauthors}%
\unskip\
\newblock
\APACrefYear{2015}.
\newblock
\APACrefbtitle {Introduction to Econometrics} {Introduction to econometrics}\
  (\PrintOrdinal{Updated third edition, global edition}\ \BEd).
\newblock
\APACaddressPublisher{{Boston Columbus Indianapolis New York San Francisco
  Hoboken Amsterdam Cape Town Dubai London}}{{Pearson}}.
\PrintBackRefs{\CurrentBib}

\bibitem [\protect \citeauthoryear {%
Storeygard%
}{%
Storeygard%
}{%
{\protect \APACyear {2016}}%
}]{%
Storeygard2016}
\APACinsertmetastar {%
Storeygard2016}%
\begin{APACrefauthors}%
Storeygard, A.%
\end{APACrefauthors}%
\unskip\
\newblock
\APACrefYearMonthDay{2016}{}{}.
\newblock
{\BBOQ}\APACrefatitle {Farther on down the Road: {{Transport}} Costs, Trade and
  Urban Growth in Sub-Saharan Africa} {Farther on down the road: {{Transport}}
  costs, trade and urban growth in sub-saharan africa}.{\BBCQ}
\newblock
\APACjournalVolNumPages{Review of Economic Studies}{83}{3}{1263--1295}.
\newblock
\begin{APACrefDOI} \doi{10/f8zzs6} \end{APACrefDOI}
\PrintBackRefs{\CurrentBib}

\bibitem [\protect \citeauthoryear {%
Verdier%
}{%
Verdier%
}{%
{\protect \APACyear {2020}}%
}]{%
verdierAverageTreatmentEffects2020}
\APACinsertmetastar {%
verdierAverageTreatmentEffects2020}%
\begin{APACrefauthors}%
Verdier, V.%
\end{APACrefauthors}%
\unskip\
\newblock
\APACrefYearMonthDay{2020}{}{}.
\newblock
{\BBOQ}\APACrefatitle {Average Treatment Effects for Stayers with Correlated
  Random Coefficient Models of Panel Data} {Average treatment effects for
  stayers with correlated random coefficient models of panel data}.{\BBCQ}
\newblock
\APACjournalVolNumPages{Journal of Applied Econometrics}{35}{7}{917--939}.
\newblock
\begin{APACrefDOI} \doi{10/gg2cp8} \end{APACrefDOI}
\PrintBackRefs{\CurrentBib}

\bibitem [\protect \citeauthoryear {%
White%
}{%
White%
}{%
{\protect \APACyear {1980}}%
}]{%
whiteHeteroskedasticityConsistentCovarianceMatrix1980}
\APACinsertmetastar {%
whiteHeteroskedasticityConsistentCovarianceMatrix1980}%
\begin{APACrefauthors}%
White, H.%
\end{APACrefauthors}%
\unskip\
\newblock
\APACrefYearMonthDay{1980}{}{}.
\newblock
{\BBOQ}\APACrefatitle {A {{Heteroskedasticity-Consistent Covariance Matrix
  Estimator}} and a {{Direct Test}} for {{Heteroskedasticity}}} {A
  {{Heteroskedasticity-Consistent Covariance Matrix Estimator}} and a {{Direct
  Test}} for {{Heteroskedasticity}}}.{\BBCQ}
\newblock

\PrintBackRefs{\CurrentBib}

\bibitem [\protect \citeauthoryear {%
Winkelmann%
}{%
Winkelmann%
}{%
{\protect \APACyear {2024}}%
}]{%
winkelmannNeglectedHeterogeneitySimpsons2024}
\APACinsertmetastar {%
winkelmannNeglectedHeterogeneitySimpsons2024}%
\begin{APACrefauthors}%
Winkelmann, R.%
\end{APACrefauthors}%
\unskip\
\newblock
\APACrefYearMonthDay{2024}{{\APACmonth{01}}}{}.
\newblock
{\BBOQ}\APACrefatitle {Neglected {{Heterogeneity}}, {{Simpson}}'s {{Paradox}},
  and the {{Anatomy}} of {{Least Squares}}} {Neglected {{Heterogeneity}},
  {{Simpson}}'s {{Paradox}}, and the {{Anatomy}} of {{Least Squares}}}.{\BBCQ}
\newblock
\APACjournalVolNumPages{Journal of Econometric Methods}{13}{1}{131--144}.
\newblock
\begin{APACrefDOI} \doi{10.1515/jem-2023-0028} \end{APACrefDOI}
\PrintBackRefs{\CurrentBib}

\bibitem [\protect \citeauthoryear {%
Wooldridge%
}{%
Wooldridge%
}{%
{\protect \APACyear {2005}}%
}]{%
wooldridgeFixedEffectsRelatedEstimators2005}
\APACinsertmetastar {%
wooldridgeFixedEffectsRelatedEstimators2005}%
\begin{APACrefauthors}%
Wooldridge, J\BPBI M.%
\end{APACrefauthors}%
\unskip\
\newblock
\APACrefYearMonthDay{2005}{{\APACmonth{05}}}{}.
\newblock
{\BBOQ}\APACrefatitle {Fixed-{{Effects}} and {{Related Estimators}} for
  {{Correlated Random-Coefficient}} and {{Treatment-Effect Panel Data Models}}}
  {Fixed-{{Effects}} and {{Related Estimators}} for {{Correlated
  Random-Coefficient}} and {{Treatment-Effect Panel Data Models}}}.{\BBCQ}
\newblock
\APACjournalVolNumPages{The Review of Economics and
  Statistics}{87}{2}{385--390}.
\newblock
\begin{APACrefDOI} \doi{10.1162/0034653053970320} \end{APACrefDOI}
\PrintBackRefs{\CurrentBib}

\end{thebibliography}

\appendix

\section{Proof of Theorem \ref{thm:consistency-PRC-theta}}\label{sec:proofs}

\begin{proof}
    \textbf{Part (i): Consistency of $\widehat{\gamma}_{n}$}.
    CITE for $\gamma$ is the coefficient estimate in a linear regression of $M_i \widetilde Y_i$ on $M_i \widetilde Z_i$, corresponding to
    \[
      M_{i} \widetilde Y_{i} = M_{i} \widetilde Z_{i}\gamma + M_{i} \widetilde U_{i}.  
    \]
    It follows that
    \begin{align*}
        \widehat{\gamma}_{n} - \gamma 
        &= \left(\frac{\sum_{i=1}^n \widetilde Z_{i}^{\prime}M_{i} \widetilde Z_{i}}{n}\right)^{-1} 
        \frac{\sum_{i=1}^n \widetilde Z_{i}^{\prime}M_{i} \widetilde U_i}{n}.
    \end{align*}
    The boundedness assumptions in the statement of the theorem,
    and the fact that the residual maker matrices $M_i$ and the linear transformation $Y_i \to \widetilde Y_i$ have bounded norm,
    imply that $E[\widetilde Z_i^\prime M_i \widetilde Z_i]$ and $E[\widetilde Z_i^\prime M_i \widetilde U_i]$ are bounded. Thus, the weak law of large numbers (WLLN) for random vectors ensures that both terms converge to their expectations.
    Because $E[\widetilde Z_i^\prime M_i \widetilde Z_i]$ is invertible by Assumption \ref{assu:rank-transformed}, it only remains to be shown that $E\left(\widetilde Z_i^\prime M_i \widetilde U_i\right)=0.$
    Assumption \ref{assu:Exogeneity-2} implies 
    $E\left(\left. \widetilde U_i \right| \widetilde X_i, \widetilde Z_i\right)=0$. 
    By the law of iterated expectations,
    $E\left(\widetilde Z_i^\prime M_i \widetilde U_i\right)=0$, since $M_i$ is a transformation of $\widetilde X_i$.

    \textbf{Part (ii): Consistency of $\widehat{\kappa}_n$}.

    The CITE for $\kappa$, cf. Definition \ref{def:DVAITE_kappa}, has the representation
    \begin{align*}
        \widehat{\kappa}_n 
        &=  \left(\sum_{i=1}^n H_{i}^\prime H_{i}\right)^{-1} 
            \sum_{i=1}^n H_{i}^\prime \widehat{\beta}_{i} \\
        &=  \left(\sum_{i=1}^n H_{i}^\prime H_{i}\right)^{-1} 
                \sum_{i=1}^n H_{i}^\prime \beta_{i} + 
            \left(\sum_{i=1}^n H_{i}^{'}H_{i}\right)^{-1} 
                \sum_{i=1}^n H_{i}^\prime \left(\widehat{\beta}_{i} - \beta_i\right).
    \end{align*}
    Because the second moments of $H$ and $\beta$ are assumed bounded in the statement of the theorem, and Assumption \ref{assu:Full-rank-H} guarantees invertibility of $E[H'H]$, the WLLN applied to the first term yields $\kappa$ by Assumption \ref{assu:Exogeneity-1}.

    If we can show that the second term converges to zero in probability, we have established consistency of $\widehat{\kappa}_n$ to $\kappa$.
    Recall the individual specific effects estimators
    \begin{align*}
        \widehat{\beta}_{i} 
        &=
        \left(\widetilde X_{i}^\prime \widetilde X_{i}\right)^{-1} 
        \widetilde X_{i}^\prime \left(\widetilde Y_{i}-\widetilde Z_{i}\widehat \gamma_n\right) \\
        &=
        \left(\widetilde X_{i}^\prime \widetilde X_{i}\right)^{-1} 
        \widetilde X_{i}^\prime 
            \left(\widetilde X_i \beta_i + 
                  \widetilde Z_{i}\left(\gamma - \widehat \gamma_n\right) + 
                  \widetilde U_i
            \right),
    \end{align*}
    where the last step uses \eqref{eq:outcome_heterogeneous_coefficients_stacked}.
    Thus, we can write
    \[
    \widehat \beta_i - \beta_i = 
    \left(\widetilde X_{i}^\prime \widetilde X_{i}\right)^{-1} 
        \widetilde X_{i}^\prime \widetilde Z_i \left(\gamma - \widehat \gamma_n\right) + 
        \left(\widetilde X_{i}^\prime \widetilde X_{i}\right)^{-1} 
        \widetilde X_{i}^\prime \widetilde U_i.
    \]

    This allows us to write
    \begin{align*}
        \widehat{\kappa}_n - \kappa 
        &= 
        \left(\sum_{i=1}^n H_{i}^\prime H_{i}\right)^{-1} \sum_{i=1}^n H_{i}^\prime \left(\widehat{\beta}_{i} - \beta_i\right) \\
        &= 
        \left(\sum_{i=1}^n H_{i}^{'}H_{i}\right)^{-1} \left[\sum_{i=1}^n H_{i}^\prime (\widetilde X_i^\prime \widetilde X_i^\prime)^{-1} \widetilde X_i^\prime \widetilde Z_i\right] \left(\gamma - \widehat \gamma_n\right) + \\
        &\phantom{=} + \left(\sum_{i=1}^n H_{i}^{'}H_{i}\right)^{-1} \sum_{i=1}^n H_{i}^\prime 
        (\widetilde X_i^\prime \widetilde X_i^\prime)^{-1} \widetilde X_i^\prime 
        \widetilde U_i \\
        &\equiv A_{1n} + A_{2n}.
    \end{align*}
    Assumption \ref{assu:switchers} ensures that $(\widetilde X_i^\prime \widetilde X_i^\prime)^{-1} \leq \frac{1}{h^2}$ uniformly over $i$, so the term can be ignored in the analysis of $A_{1n}$ and $A_{2n}$. 
    Applying Cauchy-Schwartz twice to each term, and using: boundedness of the second moments of $H$; boundedness of the fourth moments of $\widetilde X$ and $\widetilde Z$; that $\widehat \gamma_n \stackrel{p}{\to} \gamma$ from part (i); and that $E[H_{i}^\prime \widetilde X_i^\prime 
    \widetilde U_i]$ from Assumption \ref{assu:Exogeneity-2}, we conclude that $A_{1n}$ and $A_{2n}$ converge in probability to zero.
\end{proof}

\end{document}